\documentclass[pre,tighten]{revtex4}

\usepackage{amssymb,amsmath}

\usepackage{graphicx}% Include figure files

\newcommand\beq{\begin{equation}}
\newcommand\eeq{\end{equation}}
\newcommand\beqa{\begin{eqnarray}}
\newcommand\eeqa{\end{eqnarray}}
\newcommand{\dd}{\text{d}}

\newcommand{\al}{\alpha}

\usepackage{color}
\newcommand{\vicentegp}[1]{{#1}}
\begin{document}

\title{Unified hydrodynamic description for driven and undriven inelastic Maxwell mixtures at low density}
\author{Nagi Khalil\footnote[1]{Electronic address: nagi.khalil@urjc.es}}
\affiliation{Escuela Superior de Ciencias Experimentales y Tecnolog\'{\i}a (ESCET) \& GISC, Universidad Rey Juan Carlos, M\'ostoles 28933, Madrid, Spain}
\author{Vicente Garz\'{o}\footnote[2]{Electronic address: vicenteg@unex.es;
URL: http://www.unex.es/eweb/fisteor/vicente/}}
\affiliation{Departamento de F\'{\i}sica and Instituto de Computaci\'on Cient\'{\i}fica Avanzada (ICCAEx), Universidad de Extremadura, E-06006 Badajoz, Spain}

\begin{abstract}
A hydrodynamic description for inelastic Maxwell mixtures driven by a stochastic bath with friction is derived. Contrary to previous works where constitutive relations for the fluxes were restricted to states near the homogeneous \emph{steady} state, here the set of Boltzmann kinetic equations is solved by means of the Chapman--Enskog \vicentegp{method} by considering a more general time-dependent reference state. Due to this choice, the transport coefficients are given in terms of the solutions of a set of nonlinear differential equations which must be in general numerically solved. The solution to these equations gives the transport coefficients in terms of the parameters of the mixture (masses, diameters, concentration, and coefficients of restitution) and the time-dependent (scaled) parameter $\xi^*$ which determines the influence of the thermostat on the system. The Navier--Stokes transport coefficients are exactly obtained in the special cases of undriven mixtures ($\xi^*=0$) and driven mixtures under steady conditions ($\xi^*=\xi_\text{st}^*$, where $\xi_\text{st}^*$ is the value of the reduced noise strength at the steady state). As a complement, the results for inelastic Maxwell models (IMM) in both undriven and driven steady states are compared against approximate results for inelastic hard spheres (IHS) [Khalil and Garz\'o, Phys. Rev. E \textbf{88}, 052201 (2013)]. While the IMM predictions for the diffusion transport coefficients show an excellent agreement with those derived for IHS, significant quantitative differences are specially found in the case of the heat flux transport coefficients.
\end{abstract}

%\draft \pacs{05.20.Dd, 45.70.Mg, 51.10.+y, 05.60.-k}

\date{\today}
\maketitle

\section{Introduction}
\label{sec1}

Linking macroscopic laws with microscopic ones is one of the main aims of Statistical Mechanics. While this issue is well understood for macroscopic fluid systems under thermodynamic equilibrium conditions (where Gibbs' formulation connects the Hamiltonian of a system with their thermodynamic properties), a general theory for out-of-equilibrium systems is still lacking. An exception is when the fluid system is dilute enough and hence, the particles collide with short-range interactions. In this case, a kinetic theory description based on a combination of Boltzmann kinetic equation and different methods of solution has been proved to be a powerful tool. In particular, the Navier--Stokes and Burnett hydrodynamic equations with explicit expressions for the transport coefficients have been derived for general potential interactions by solving the Boltzmann kinetic equation by means of the Chapman--Enskog method \cite{CC70}. This perturbative method is based on the expansion of the distribution function around a chosen \emph{reference} state, a state where the system keeps close to.

The Chapman--Enskog method has been mainly employed to solve the Boltzmann equation for ordinary or molecular gases (namely when the collisions among particles are elastic). In this case, the solution to the Boltzmann equation in the absence of spatial gradients (zeroth-order approximation) is given by the local version of the Maxwell-Boltzmann velocity distribution function, namely, the distribution function  obtained from the Maxwell--Boltzmann distribution by replacing temperature, density and flow velocity with their actual nonequilibrium values. Since a well-known feature of the equilibrium state is that the gas evolves spontaneously towards it after a few collisions per particle \cite{MGB18,K19} (regardless of the initial preparation of the system), the election of the above reference state is well justified for ordinary gases. However, when the number of particles \cite{GMSBT08,MGSBT08,GMSBT09}, the linear momentum \cite{EGLM19}, and/or the kinetic energy \cite{G03,BP04,G19} are not conserved in collisions, then the situation becomes more cumbersome and the choice of a proper reference state is not simple nor even unique.

A natural question to ask in all the above situations is: What is the appropriate reference state to be used in a perturbative method like the Chapman--Enskog method? As said before, for ordinary fluids close to thermal equilibrium, a good choice is the local Maxwell--Boltzmann distribution function. However,  in the case of systems inherently out of equilibrium such as granular gases (a gas constituted by macroscopic particles that undergo inelastic collisions), the Maxwell--Boltzmann distribution is not a solution of the homogeneous (inelastic) Boltzmann equation and hence, we have to look for another reference distribution function. In particular, for freely cooling granular gases, the zeroth-order approximation in the Chapmann-Enskog expansion is the local version of the so-called homogeneous cooling state, namely, a homogeneous state where the granular temperature monotonically decays in time \cite{G03,BP04,K18,G19}. The homogeneous cooling state has been widely used as the reference state in the Chapman--Enskog method to obtain not only the general form of the hydrodynamic equations, but also to explicitly determine the expressions of the Navier--Stokes \cite{BDKS98,GD99a} and Burnett \cite{KGS14} transport coefficients. Although this reference state is a time-dependent state (since the temperature decreases in time due to the collisional cooling), the resulting hydrodynamic equations describe reasonably well for not strong values of inelasticity the transport properties of unsteady and steady states eventually reached by the system when energy is injected through the boundaries \cite{BRM00,BRM01,BKR09,VSG13}. However, when the energy input is done globally \cite{AD06,SGS05} or by means of a vibrating plate \cite{OU05,RPGRSCM11,GSVP11,CMS12,BRS13,BBMG15}, it is more convenient to take a time-dependent reference state different from the conventional homogeneous cooling state.

Beyond the homogeneous cooling state, another type of reference states can be chosen when, for instance, the granular gas is strongly sheared \cite{SGD04,L06,G06a,VSG13} or subjected to strong temperature gradients \cite{BKD11,BKD12,K16}. Another relevant situation is when the granular gas is driven by the action of an external driving force or thermostat \cite{EM90}. This is the usual way to drive a granular gas in computer simulations \cite{PLMPV98,PLMV99,PBL02,PTNE02,PEU02,FAZ09,SVGP10,SVCP10,VAZ11,GSVP11,SSW13}. In the case of spatially homogeneous situations, when the energy injected by the thermostat is exactly compensated for by the energy lost by collisions, a nonequilibrium \emph{steady} state is reached, a state analogous to the equilibrium state of molecular gases. However, the above steady state could not be a good choice for the reference state in the Chapman--Enskog solution, since a local election of the hydrodynamic variables induces a collisional cooling that, in general, cannot be exactly compensated for by the energy injected in the system by the thermostat \cite{GMT12}. This means that the dynamics close to the steady state requires a time-dependent reference homogeneous solution to the kinetic equation. This is a subtle and important point that must be taken into account when one attempts to obtain the transport properties.

The Navier--Stokes transport coefficients of driven granular gases modeled as inelastic hard spheres (IHS) have been recently obtained for mono \cite{GChV13,GChV13bis,GGG19a} and multicomponent \cite{KG13,KG18,KG19} systems. In the above papers, the gas is driven by a stochastic bath with friction. However, there are two important limitations in the above works. First, although the reference state is a time-dependent distribution, the explicit forms of the transport coefficients were derived by assuming steady state conditions, namely, when there is an exact balance between the energy input and the energy dissipated by collisions. This allowed us to get analytical expressions for the Navier--Stokes transport coefficients. Second, due to the mathematical complexity of the Boltzmann collision operator, the results were approximately achieved by considering the leading terms in a Sonine polynomial expansion. This second limitation can be overcome by considering the so-called inelastic Maxwell models (IMM) \cite{BK00,BCG00,EB02a,EB02b,BK03}: a model where the collision rate of two particles about to collide is assumed to be independent of their relative velocity. As in the case of the conventional Maxwell molecules \cite{TM80}, the above collisional simplification allows us to obtain the \emph{exact} forms of the velocity moments of the velocity distribution functions \cite{SG95,GS07} without their explicit knowledge.

The main objective of this work is to provide a closed Navier--Stokes hydrodynamic description of driven granular mixtures. Our starting point is the set of kinetic Boltzmann equations for IMM that is solved by means of the Chapman--Enskog expansion around a time-dependent reference state which can be arbitrarily far away from the homogeneous steady state. This type of description differs from the one previously reported \cite{KG13} where the expressions of the transport coefficients were restricted to states \emph{close} to the homogeneous steady state. In the present work, the choice of a general time-dependent reference state provides a general hydrodynamic description where, for instance, we can find regions in the system where the transport coefficients are very close to those obtained for undriven granular mixtures together with other regions where the dynamics is dominated by the effect of the bath or thermostat. As an intermediate situation, an exact balance between dissipation in collisions and energy injected by the thermostat (steady state conditions) can be seen as well. In this context, the present theory include all previous ones \cite{GD02,KG13}, which are recovered taking the appropriate limits.

Since the determination of the complete set of transport coefficients for driven granular mixtures requires long and complex calculations, here we consider IMM instead of IHS. This makes the presentation simpler as well as the achieved results exact, without the need of additional and sometimes uncontrolled approximations. In any case, the methodology employed here for IMM can be adapted to IHS for the determination of its corresponding transport coefficients; most of the present results being intuitively extrapolated to other models of driven granular gases.

In contrast to previous derivations for undriven \cite{GD02,GA05} and driven \cite{KG13} granular mixtures, the transport coefficients associated with the mass flux, the pressure tensor, and the heat flux are given in terms of the solution of a set of nonlinear coupled differential equations. These differential equations involve the derivatives of the (scaled) transport coefficients with respect to the scaled parameter of the thermostat $\xi^*$. The above differential equations can be analytically solved in two cases: (i) undriven granular mixtures ($\xi^*=0$; whose results were already reported in Ref.\ \cite{GA05}) and (ii) driven mixtures in steady state conditions ($\xi^*=\xi_\text{st}^*$, where $\xi_\text{st}^*$ is the value of the reduced noise strength at the steady state). Beyond these limit cases, the transport coefficients are obtained via a numerical integration of the above set of differential equations. Apart from the usual transport coefficients, our results show that the first-order contributions $T_i^{(1)}$ to the partial temperatures are different from zero. This new contribution (which is also present in dense mixtures \cite{KS79a,GGG19b}) to the breakdown of energy equipartition was neglected in previous works \cite{KG13,KG18} on driven mixtures. Since this contribution is proportional to the divergence of the flow velocity, it is involved then in the evaluation of the first-order contribution $\zeta_U$ to the cooling rate. Our results show that the magnitude of $T_i^{(1)}$ (see for instance, Fig.\ \ref{fig10})  can be significant in some regions of the parameter space of the system. The fact that $T_i^{(1)}\neq 0$ contrasts with the results for undriven granular mixtures \cite{GD02,GA05} since this coefficient vanishes in the low-density limit.

The organization of the paper is as follows. In section \ref{sec2} we introduce the model as well as the kinetic and hydrodynamic descriptions. The reference time-dependent state is analyzed in section \ref{sec3} where it is shown that this state reduces to both the homogeneous cooling state and the homogeneous steady state in their corresponding limits. The Chapman--Enskog method is briefly described in section \ref{sec4} while the kinetic equation verifying the first-order distribution function is provided in section \ref{sec5}. Technical details on the determination of the Navier--Stokes transport coefficients as well as the first-order contributions to the partial temperatures are relegated to Appendices \ref{appA} and \ref{appB}. As said before, the transport coefficients are given in terms of the solution of a set of nonlinear coupled differential equations. These equations are solved for some representative cases, showing the dependence of the transport coefficients on the parameters of the system. As a complement, a comparison with the results obtained in previous works for IHS \cite{KG13,KG18} in steady state conditions is also addressed in section \ref{sec6}. The paper ends in section \ref{sec7} with a brief discussion of the results reported along the text.

%%%%%%%%%%%%%%%%%%%%%%%%%%%%%%%%%%%%%%%%%%%%%%%%%%%%%%%%%%%%%%%%%%%%%%%%%%%%%%%%%%%%%%%%%%%%%%%%%%%%%%%%%%%%%%%%%%%%%%%%%%%%%%%%%%%%%%

\section{Boltzmann kinetic theory and hydrodynamics}
\label{sec2}

\subsection{Model and kinetic description}

Consider a granular binary mixture modeled as a binary mixture of inelastic Maxwell gases at low density. The Boltzmann equation for IMM \cite{BK00,BCG00,EB02a,EB02b,BK03} can be obtained from the Boltzmann equation for IHS by replacing the  rate for collisions between particles of components $i$ and $j$ by an average  velocity-independent collision rate, which is proportional to the square root of the ``granular''  temperature $T$ (defined later). With this simplification, the velocity distribution function $f_i({\bf r}, {\bf v};t)$ of a particle of component $i$ $(i=1, 2)$ with position $\bf r$ and velocity $\bf v$ at time $t$ satisfies the following set of nonlinear Boltzmann kinetic equations:
\beq
\label{2.1}
\partial_{t}f_i+\mathbf{v}\cdot \nabla f_i+{\mathcal F}_if_i=\sum_{j=1}^2\; J_{ij}[\mathbf{v}|f_i,f_j],
\eeq
where the Boltzmann collision operator $J_{ij}[f_i,f_j]$ for IMM in $d$ dimensions is \cite{G19}
\begin{equation}
\label{2.2}
J_{ij}[{\bf v}_{1}|f_{i},f_{j}] =\frac{\nu_{ij}}{n_j\Omega_d}\int \dd{\bf v}_{2}\int \dd\widehat{\boldsymbol{\sigma }}\left[ \alpha_{ij}^{-1}f_i(\mathbf{r},\mathbf{v}_1',t)f_{j}(\mathbf{r},\mathbf{v}_2',t)-f_{i}(\mathbf{r},\mathbf{v}_1,t)
f_{j}(\mathbf{r},\mathbf{v}_2,t)\right].
\end{equation}
Here,
\beq
\label{2.3}
n_{i}=\int \dd\mathbf{v}\;f_{i}(\mathbf{v})
\eeq
is the number density of component $i$, $\nu_{ij}$ is an effective collision frequency (to be chosen later) for collisions  of type $i$-$j$,  $\Omega_d=2\pi^{d/2}/\Gamma(d/2)$ is the total solid angle in $d$ dimensions, and $\alpha_{ij}=\alpha_{ji}\leq 1$ refers to the constant coefficient of normal restitution for collisions between particles of component $i$ with $j$. Although negative values of $\alpha_{ij}$ can be considered \cite{K18}, we restrict ourselves in this work to positive values of $\alpha_{ij}$. In Eq.\ \eqref{2.2}, the relationship between the pre-collisional $\{{\bf v}_{1}', {\bf v}_{2}'\}$ and post-collisional $\{{\bf v}_{1},{\bf v}_{2}\}$ velocities is:
\begin{equation}
\label{2.4}
{\bf v}_{1}'={\bf v}_{1}-\mu_{ji}\left( 1+\alpha_{ij}
^{-1}\right)(\widehat{\boldsymbol{\sigma}}\cdot {\bf g}_{12})\widehat{\boldsymbol
{\sigma}},
\quad {\bf v}_{2}'={\bf v}_{2}+\mu_{ij}\left(
1+\alpha_{ij}^{-1}\right) (\widehat{\boldsymbol{\sigma}}\cdot {\bf
g}_{12})\widehat{\boldsymbol{\sigma}}\;,
\end{equation}
where ${\bf g}_{12}={\bf v}_1-{\bf v}_2$ is the relative velocity of the colliding pair,
$\widehat{\boldsymbol{\sigma}}$ is a unit vector directed along the centers of the two colliding
spheres, $\mu_{ij}=m_i/(m_i+m_j)$, and $m_i$ is the mass of component $i$. The collision rules \eqref{2.4} conserve the number of particles of each component and the total linear momentum. However, the total kinetic energy of the colliding pair is reduced by a factor $1-\alpha_{ij}^2$ after the collision, hence $\alpha_{ij}=1$ and $\alpha_{ij}=0$ correspond to the elastic and completely inelastic limits, respectively.

The operator ${\mathcal F}_i$ in the Boltzmann equation \eqref{2.1} accounts for the effect of an external force (or thermostat) on particles of component $i$. The external force has two contributions: (i) a frictional or drag force proportional to the relative velocity $\mathbf{v}-\mathbf{U}_g$ ($\mathbf{U}_g$ being the known flow velocity of the background or interstitial gas), and (ii) a stochastic force with the form of a Gaussian white noise \cite{WM96}. Thus, the operator ${\mathcal F}_i$ has the form \cite{KG13}
\beq
\label{2.5}
\mathcal{F}_if_i=-\frac{\gamma_\text{b}}{m_i^{\beta}}
\frac{\partial}{\partial\mathbf{v}}\cdot \left(\mathbf{v}-\mathbf{U}_g\right)f_i-\frac{1}{2}\frac{\xi_\text{b}^2}{m_i^{\lambda}}\frac{\partial^2 f_i}{\partial v^2},
\eeq
where $\gamma_\text{b}$ is the drag (or friction) coefficient and $\xi_\text{b}^2$ represents the strength of the correlation in the Gaussian white noise. Moreover, $\beta$ and $\lambda$ are \emph{arbitrary} constants of the driven model.

As widely discussed in Ref.\ \cite{KG13}, the model \eqref{2.5} is a generalization of previous driven models since it coincides with them for specific values of $\beta$ and $\lambda$. In particular, when $\gamma_\text{b}=0$ and $\lambda=0$, our thermostat reduces to the stochastic thermostat employed in several papers \cite{HBB00,BT02} for conducting numerical simulations in granular mixtures. The choice $\beta=1$ and $\lambda=2$ leads to the conventional Fokker--Planck model for molecular mixtures \cite{H03}. This latter sort of thermostat has been extensively employed \cite{PLMPV98,PLMV99,PBL02,SVGP10,SVCP10,GSVP11}, specially when studying granular Brownian motion. As a third possibility, the choice $\xi_\text{b}^2=2 \gamma_\text{b} T_\text{b}$ ($T_\text{b}$ being the background or bath temperature), $\beta=0$, and $\lambda=1$ implements a force $\mathcal{F}_i$ quite similar to the fluid-solid interaction force that models the effect of the viscous gas on monodisperse solid particles \cite{KIB14,HTG17}. It is also interesting to remark that the term \eqref{2.5} has been derived from the formalism describing the general interaction between particles of the component $i$ with a thermal bath \cite{K81,KG14}. The main assumption for deriving \eqref{2.5} is that the action of the bath on component $i$ depends only on its velocity distribution $f_i$. More details on the driven model \eqref{2.5} can be found in Refs.\ \cite{KG13,KG14}.

Taking into account the form \eqref{2.5} of the forcing term $\mathcal{F}_if_i$, the Boltzmann equation \eqref{2.1} becomes
\beq
\label{2.7}
\partial_{t}f_i+\mathbf{v}\cdot \nabla f_i-\frac{\gamma_\text{b}}{m_i^{\beta}}\Delta \mathbf{U} \cdot \frac{\partial}{\partial\mathbf{v}}f_i-\frac{\gamma_\text{b}}{m_i^{\beta}} \frac{\partial}{\partial\mathbf{v}}\cdot \mathbf{V} f_i -\frac{1}{2}\frac{\xi_\text{b}^2}{m_i^{\lambda}}\frac{\partial^2}{\partial v^2}f_i=\sum_{j=1}^2\; J_{ij}[\mathbf v|f_i,f_j],
\eeq
where $\Delta \mathbf{U}=\mathbf{U}-\mathbf{U}_g$, $\mathbf{V}=\mathbf{v}-\mathbf{U}$ is the peculiar velocity, and
\beq
\label{2.6}
\mathbf{U}=\rho^{-1}\sum_{i=1}^2\int\dd \mathbf{ v}m_{i}\mathbf{v}f_{i}(\mathbf{v})
\end{equation}
is the mean flow velocity of grains. In Eq.\ \eqref{2.6}, $\rho=\sum_i \rho_i$ is the total mass density and $\rho_i=m_i n_i$ is the mass density of component $i$.

The Boltzmann collision operators have the following properties:
\begin{equation}
\int \dd\mathbf{ v}\;J_{ij}[\mathbf{ v}|f_{i},f_{j}]=0, \quad  \sum_{i=1}^2\sum_{j=1}^2m_i\int \dd\mathbf{ v}\;\mathbf{v}\; J_{ij}[{\bf v}|f_{i},f_{j}]=0, \label{2.8}
\end{equation}
\begin{equation}
\sum_{i=1}^2\sum_{j=1}^2m_i\int \dd\mathbf{ v}\; V^{2}\; J_{ij}[\mathbf{v}|f_{i},f_{j}]\equiv-d nT\zeta.  \label{2.10}
\end{equation}
The last equality defines the total ``cooling rate'' $\zeta$ due to inelastic collisions among all species,
\beq
\label{2.11}
T=\frac{1}{n}\sum_{i=1}^2\int \dd\mathbf{v}\frac{m_{i}}{d}V^{2}f_{i}(\mathbf{ v})\;
\eeq
is the granular temperature, and $ n=n_{1}+n_{2}$ the total number density. Moreover, an interesting quantity at a kinetic level is the partial kinetic temperature $T_i$ defined as
\begin{equation}
\label{2.12}
T_i=\frac{m_{i}}{d n_i}\int\; \dd\mathbf{ v}\;V^{2}f_{i}(\mathbf{ v}).
\end{equation}
The granular temperature $T$ can also be written as
\beq
\label{2.12.1}
T=\sum_{i=1}^2\, x_i T_i,
\eeq
where $x_i=n_i/n$ is the mole fraction of species $i$. We can introduce the partial cooling rates $\zeta_i$ associated with the partial temperatures $T_i$ as
\begin{equation}
\label{2.13}
\zeta_i=\sum_{j=1}^2\zeta_{ij}=-\frac{m_i}{dn_iT_i}\sum_{j=1}^2\int \dd\mathbf{ v}\; V^{2}J_{ij}[{\bf v}|f_{i},f_{j}],
\end{equation}
where $\zeta_{ij}$ are defined thought the second equality. As for the granular temperature, the total cooling rate $\zeta$ can be written as
\begin{equation}
\label{2.14}
\zeta=T^{-1}\sum_{i=1}^2\;x_iT_i\zeta_i.
\end{equation}

As happens for elastic Maxwell molecules \cite{TM80}, the collisional moments of the Boltzmann operator $J_{ij}[f_i,f_j]$ for IMM can be \emph{exactly} computed without the knowledge of the velocity distributions $f_i$ and $f_j$ \cite{BC02,GS07}. In particular, the quantities $\zeta_{ij}$ (which define the cooling rate $\zeta$) are given by \cite{G03bis}
\begin{equation}
\label{2.15}
\zeta_{ij}=\frac{2\nu_{ij}}{d}\mu_{ji}(1+\alpha _{ij})\left[1-\frac{\mu_{ji}}{2}(1+\alpha_{ij})
\frac{\theta_i+\theta_j}{\theta_j}+\frac{\mu_{ji}(1+\alpha_{ij})-1}{d\rho_jp_i}
{\bf j}_i\cdot {\bf j}_j\right],
\end{equation}
where $\theta_i= m_i T/\overline{m}T_i$, $\overline{m}=m_1m_2/(m_1+m_2)$ is the reduced mass, $p_i=n_i T_i$ is the partial pressure of component $i$ and
\begin{equation}
\label{2.16}
\mathbf j_i=m_i\int \dd\mathbf v \ \mathbf V\; f_i(\mathbf v)
\end{equation}
is the mass flux of component $i$ relative to the local flow. According to Eq.\ \eqref{2.12.1}, the hydrostatic pressure is $p=\sum_i p_i=n T$.

In order to fully define the model we still have to choose the collision frequencies $\nu_{ij}$ of Eq. \ \eqref{2.2}. As in previous works on IMM \cite{G03bis,GA05}, $\nu_{ij}$ are chosen so that the partial cooling rates $\zeta_{ij}$ coincide with that of IHS in the so-called homogeneous cooling state \cite{G19}. With this choice, $\nu_{ij}$ is defined as
\beq
\label{2.17}
\nu_{ij}=\frac{\Omega_d}{\sqrt{\pi}} x_j\left(\frac{\sigma_{ij}}{\sigma_{12}}\right)^{d-1} \left(\frac{\theta_i+\theta_j}{\theta_i\theta_j}\right)^{1/2}\nu_0,
\end{equation}
where $\sigma_{ij}=(\sigma_i+\sigma_j)/2$, $\sigma_i$ is the diameter of particles of component $i$, and
\begin{equation}
\label{2.18}
\nu_0=n\sigma_{12}^{d-1}\sqrt{\frac{2T}{\overline{m}}}
\end{equation}
is an effective collision frequency. Upon deriving Eq.\ \eqref{2.17} use has been made of the fact that the mass flux $\mathbf{j}_i=\mathbf{0}$ in the homogeneous cooling state.

\begin{table}[!h]
  \centering
  \begin{tabular}{c|c|c|c}
    Symbol & Name & Definition & Equations \\ \hline \hline
    $\Omega_d$ & Solid angle in $d$ dimensions &$2\pi^{d/2}/\Gamma(d/2)$ & \eqref{2.2} \\
    $\sigma_{ij}$ & &$(\sigma_i+\sigma_j)/2$  & \eqref{2.17} \\
    $\mu_{ij}$ & & $m_i/(m_i+m_j)$  & \eqref{2.4} \\
    $\overline m$ & Reduced mass & $m_1m_2/(m_1+m_2)$ & After \eqref{2.15} \\
    $M_i$ & & $m_i/\overline m$  & \eqref{3.6} \\
    $n_i$ & Number density of component $i$ & $\int\ d\mathbf v f_i  $ & \eqref{2.2} \\
    $n$ & Total number density & $p/T$ &  \eqref{2.11} \\
    $x_1$ & Mole fraction of component 1 & $n_1 T/p$ & \eqref{2.12.1}\\
    $x_2$ & Mole fraction of component 2 & $1-x_1$ & \eqref{2.12.1} \\
    $\rho_1$ & Mass density of component 1 & $m_1 p x_1/T$ & After \eqref{2.6} \\
    $\rho_2$ & Mass density of component 2 & $m_2 p(1-x_1)/T$ & After \eqref{2.6} \\
    $ \rho$ & Total mass density & $p\left[m_1x_1+m_2(1-x_1)\right]/T$ & \eqref{2.6}\\
    $\chi_1$ & Zeroth-order temperature ratio of component 1 & $T_1/T$ & \eqref{3.2}\\
    $\chi_2$ & Zeroth-order temperature ratio of component 2 & $(1-x_1 \chi_1)/(1-x_1)$ & \eqref{3.2}\\
    $\theta_1$ & & $M_1/\chi_1$  & \eqref{2.15}\\
    $\theta_2$ & & $M_2(1-x_1)/(1-x_1\chi_1)$ & \eqref{2.15}\\
    %$\theta $ & $\theta_1/\theta_2$ & $M_1(1-x_1\chi_1)/[M_2\chi_1(1-x_1)]$ & \\
    $v_0$ & Thermal speed & $v_0=\sqrt{2T/\overline m}$ & \eqref{3.4}\\
    $\nu_0$ & Effective collision frequency & $n\sigma_{12}^{d-1}v_0$ & \eqref{2.17}\\
    $\nu_{ij}$ & Collision frequencies & $(\Omega_d/\sqrt{\pi})(\sigma_{ij}/\sigma_{12})^{d-1}x_j \sqrt{(\theta_i+\theta_j)/\theta_i\theta_j}\nu_0$ & \eqref{2.2}\\
    $\omega^*$ & Dimensionless drift &$(\gamma_\text{b}/\overline{m}^\beta) \left(\overline{m}^\lambda/2 \xi_\text{b}^2\right)^{1/3} \left(p\sigma_{12}^{d-1}/T\right)^{-2/3}$ & \eqref{3.4}\\
    $\xi^*$ & Dimensionless noise & $\xi_\text{b}^2/(\nu_0T\overline m^{\lambda-1})$ & \eqref{3.4} \\
    $\zeta_i$ & Partial cooling rate of component $i$ & Eq.\ \eqref{2.13} & Before \eqref{2.13}
%    $\zeta_i'$ & &$\sqrt{M_i/\chi_i}\zeta_i/\nu_0$ &
  \end{tabular}
  \caption{Definitions of some of the quantities appearing along the text. The equations where this quantities appear for the first time are provided by the last column.}
  \label{tab:1}
\end{table}

Several observations are in order. On the one hand, not all quantities are independent since, for instance, we have $x_1+x_2=1$, $T=x_1T_1+x_2T_2$, and $\mathbf j_1=-\mathbf j_2$. In addition, as will be shown later, the partial temperatures $T_i$, and hence $\theta_i$, have nonzero contributions in the Navier--Stokes domain (first-order in spatial gradients), namely $T_i=T_i^{(0)}+T_i^{(1)}$ with $T_i^{(1)}\ne 0$ in general. These contributions are proportional to the divergence of the flow velocity, namely, $T_i^{(1)} \propto \nabla\cdot \mathbf{U}$. Thus, for the sake of simplicity, the partial temperatures defining the collision frequencies $\nu_{ij}$ in Eq.\ \eqref{2.17} are taken to be of order zero in spatial gradients (i.e., $\theta_i= m_i T/\overline{m}T_i^{(0)}$). Table \ref{tab:1} collects most of the definitions employed along the paper.

\subsection{Hydrodynamic description}

The hydrodynamic balance equations for $n_i$, $\mathbf{U}$, and $T$ can be easily derived by multiplying the set of Boltzmann equations \eqref{2.7} by $m_i$, $m_i \mathbf{v}$, and $\frac{1}{2}m_i V^2$, respectively, integrating over velocity, and taking into account the properties \eqref{2.8}--\eqref{2.10} of the operator $J_{ij}[f_i,f_j]$. Then, the corresponding hydrodynamic equations for the mole fraction $x_1=n_1/n$, the hydrostatic pressure $p$, the temperature $T$, and the $d$ components of the local flow velocity $\mathbf{U}$ can be easily obtained:
\begin{eqnarray}
\label{2.19}
&& D_tx_1+\frac{\rho}{n^2m_1m_2}\nabla\cdot \mathbf j_1=0, \\
\label{2.20}
&& D_{t}\mathbf{U}+\rho ^{-1}\nabla \cdot\mathsf{P}=-\frac{\gamma_\text{b}}{\rho}\left(\Delta \mathbf{U}\sum_{i=1}^2\frac{\rho_i}{m_i^\beta} +\sum_{i=1}^2\frac{\mathbf{ j}_{i}}{m_{i}^\beta}\right), \\
\label{2.21}
&& D_{t}T-\frac{T}{n}\sum_{i=1}^2\frac{\nabla \cdot \mathbf{ j}_{i}}{m_{i}}+\frac{2}{dn} \left( \nabla \cdot \mathbf{ q}+\mathsf{P}:\nabla \mathbf{U}\right)=-\zeta \,T -\frac{2 \gamma_\text{b}}{d n} \sum_{i=1}^2\frac{ \Delta \mathbf{U} \cdot\mathbf{ j}_{i}}{m_{i}^\beta}-2\gamma_\text{b}\sum_{i=1}^2 \frac{x_i T_i}{m_i^\beta}+\frac{\xi_\text{b}^2}{n}\sum_{i=1}^2\frac{\rho_i}{m_i^\lambda}, \\
\label{2.22}
& & D_t p+ p \nabla \cdot \mathbf{U}+\frac{2}{d}\left( \nabla \cdot \mathbf{ q}+\mathsf{P}:\nabla \mathbf{U}\right)=-\zeta \,p-\frac{2\gamma_\text{b}}{d}\sum_{i=1}^2 \frac{\Delta \mathbf{U}\cdot \mathbf{j}_i}{m_i^\beta}-2\gamma_\text{b}\frac{p}{T}\sum_{i=1}^2 \frac{x_i T_i}{m_i^\beta}+\xi_\text{b}^2\sum_{i=1}^2\frac{\rho_i}{m_i^\lambda}.
\end{eqnarray}
In Eqs.\ \eqref{2.19}--\eqref{2.22}, $D_t=\partial_t+\mathbf U\cdot \nabla$ is the material derivative, $\mathbf{j}_i$ is defined by Eq.\ \eqref{2.16},
\begin{equation}
\label{2.23}
\mathsf P=\sum_{i=1}^2m_i\int \dd\mathbf v\ \mathbf V\mathbf V f_i(\mathbf v)
\end{equation}
is the total pressure tensor,
\begin{equation}
\label{2.24}
\mathbf q=\sum_{i=1}^2\frac{m_i}{2}\int \dd\mathbf v\ V^2\mathbf V f_i(\mathbf v)
\end{equation}
is the total heat flux, and the cooling rate  $\zeta$ is defined by Eq.\ \eqref{2.10}. Note that the balance equations \eqref{2.19}--\eqref{2.22} apply for both interaction models IHS and IMM. The difference between both models is unveiled when the explicit forms of the Boltzmann collision operators are accounted for in the evaluation of the transport coefficients and the cooling rate.

\vicentegp{Note that the balance equations \eqref{2.19}--\eqref{2.22} [which are a direct consequence of the properties \eqref{2.8}--\eqref{2.10} of the Boltzmann collision operators] are local versions of the (macroscopic) conservation laws. These hydrodynamic laws could in principle be derived independently of the kinetic theory viewpoint by invoking symmetry considerations. On the other hand, a kinetic description provides a clear bridge between microscopic (dynamics of two grains) and macroscopic (hydrodynamics fields) descriptions that make easier the derivation of hydrodynamic equations}. 

The balance equations \eqref{2.19}--\eqref{2.22} become a closed set of differential equations for the hydrodynamic fields once the irreversible fluxes and the cooling rate are expressed in terms of the hydrodynamic fields. These relations are the so-called constitutive equations. This goal can be achieved by solving the Boltzmann equations by means of the well-known Chapman--Enskog method \cite{CC70} adapted to driven granular mixtures.

The determination of the Navier-Stokes transport coefficients for IHS was accomplished in Refs.\ \cite{KG13,KG18}. Nevertheless, due to the mathematical difficulties of the problem, only \emph{steady} state conditions were considered. In this paper, we revisit this problem in the case of IMM (where all the results are exact) but for arbitrary \emph{unsteady} conditions. This allows us to determine the transport coefficients not only in the steady state but also for more general physical conditions. Since the characterization of the time-dependent homogeneous state is essential for deriving the Navier--Stokes hydrodynamic equations, before considering inhomogeneous situations we will study first the homogeneous states. This will be carried out in the following section.

\section{Homogeneous time-dependent states}
\label{sec3}

As extensively discussed in Refs. \cite{GMT12,K18} for monocomponent granular gases, two separate stages can be identified in the dynamical evolution of a system from any initial condition. A \emph{fast} first stage (for times of the order of the mean free time) can be identified where the evolution of the gas clearly depends on the initial preparation of the system. Then, a second \emph{slow} stage can be observed where the evolution of the gas is completely determined by the time evolution of the hydrodynamic fields. While the first stage defines the so-called \emph{kinetic} regime, the second one refers to the so-called \emph{hydrodynamic} regime. Here, we are interested in the hydrodynamic regime. \vicentegp{A more systematic description of the system in the fast stage regime is, in principle, possible \cite{somamimoretr15,paklgr18,grklpa20} but beyond the scope of the present work}.

In homogeneous states, the concentration $x_1$ is constant, the pressure $p$ and temperature $T$ are spatially uniform, and with an appropriate choice of the reference frame $\mathbf U_g=\mathbf U=\mathbf 0$. Under these conditions, the set of Boltzmann equations \eqref{2.7} becomes
\begin{equation}
\label{3.1}
\partial_{t}f_i-\frac{\gamma_\text{b}}{m_i^{\beta}} \frac{\partial}{\partial\mathbf{v}}\cdot \mathbf{v} f_i-\frac{1}{2}\frac{\xi_\text{b}^2}{m_i^{\lambda}}\frac{\partial^2 f_i}{\partial v^2}=\sum_{j=1}^2\; J_{ij}[\mathbf{v}|f_i,f_i].
\end{equation}
The balance equations \eqref{2.19}--\eqref{2.22} for homogeneous states simply reduce to $\partial_t x_1=\partial_t U_i=0$ and $T^{-1}\partial_t T=p^{-1}\partial_t p=-\Lambda$, where
\begin{equation}
\label{3.2}
\Lambda = \zeta+2\gamma_\text{b}\sum_{i=1}^2 \frac{x_i \chi_i}{m_i^\beta}-\frac{\xi_\text{b}^2}{p}\sum_{i=1}^2\frac{\rho_i}{m_i^\lambda},
\end{equation}
and $\chi_i=T_i/T$. Since the time dependence of the distribution functions enters through $T$ and $p$ in the hydrodynamic regime, we have the identity $\partial_t f_i=-\Lambda \left(T\partial_T+p\partial_p\right)f_i$ and the Boltzmann equation \eqref{3.1} reads
\begin{equation}
\label{3.3}
-\Lambda\left(T\partial_T+p\partial_p\right)f_i-\frac{\gamma_\text{b}}{m_i^{\beta}} \frac{\partial}{\partial\mathbf{v}}\cdot \mathbf{v} f_i-\frac{1}{2}\frac{\xi_\text{b}^2}{m_i^{\lambda}}\frac{\partial^2 f_i}{\partial v^2}=\sum_{j=1}^2\; J_{ij}[\mathbf{v}|f_i,f_i].
\end{equation}

As discussed in Ref.\ \cite{KG13}, the solutions to the set of coupled Boltzmann equations \eqref{3.3} have the scaling forms
\begin{equation}
\label{3.4}
f_i(\mathbf v,t)=x_i\frac{p(t)}{T(t)}v_0(t)^{-d}\varphi_i(x_1,\mathbf c,\omega^*,\xi^*),
\end{equation}
where $v_0(t)=\sqrt{2T(t)/\overline m}$ is the thermal speed, $\mathbf c=\mathbf v/v_0$ is the scaled velocity, and we have introduced the following dimensionless thermostat parameters:
\beq
\label{3.5}
\omega^*= \frac{\gamma_\text{b}}{\overline{m}^\beta} \left(\frac{\overline{m}^\lambda}{2 \xi_\text{b}^2}\right)^{1/3} \left(\frac{p\sigma_{12}^{d-1}}{T}\right)^{-2/3}, \quad \xi^*= \frac{\xi_\text{b}^2}{\nu_0 T\overline m^{\lambda-1}}.
\eeq
Since the total number density $n=p/T$ is independent of time, then $\omega^*$ does not depend on time. Equation \eqref{3.4} reveals that the time dependence of the scaled distribution $\varphi_i$ is encoded through the dimensionless velocity $\mathbf{c}$ and the (reduced) noise strength $\xi^*$.

In terms of the above dimensionless parameters, Eq.\ \eqref{3.2} can be rewritten as
\beq
\label{3.6}
\Lambda^* \left(\frac{1}{2}\frac{\partial}{\partial\mathbf{ c}}\cdot \mathbf{ c}\varphi_i+\frac{3}{2}\xi^*\frac{\partial \varphi_i}{\partial \xi^*}\right)-\frac{\omega^* \xi^{*1/3}}{M_i^\beta}\frac{\partial}{\partial \mathbf{c}}\cdot \mathbf{ c} \varphi_i-\frac{1}{4}\frac{\xi^*}{M_i^\lambda} \frac{\partial^2 \varphi_i}{\partial c^2}=\sum_{i=1}^2 J_{ij}^*[\varphi_i,\varphi_j],
\eeq
where $M_i=m_i/\overline m$, $\Lambda^*=\Lambda/\nu_0$, and
\beqa
\label{3.7}
J_{ij}^*[\mathbf{c}|\varphi_{i},\varphi_{j}]&=& \frac{J_{ij}[\mathbf{v}|f_{i},f_{j}]}
{n_{i}v_0^{-d}\nu_0}\nonumber\\
&=&\frac{\nu_{ij}^*}{\Omega_d}\int
\dd\mathbf{c}_{2}\int \dd\widehat{\boldsymbol {\sigma}}\left[ \alpha_{ij}^{-1}\varphi_{i}(\mathbf{c}_{1}^{\prime
},t)\varphi_{j}(\mathbf{c}_{2}^{\prime},t)-\varphi_{i}
(\mathbf{c}_{1},t)\varphi_{j}(\mathbf{c}_{2},t)\right],
\eeqa
with $\nu_{ij}^*=\nu_{ij}/\nu_0$. As for IHS, the solution to Eq.\ \eqref{3.6} is not known. Nevertheless, the form of the Boltzmann operator for IMM allows us to determine exactly its velocity moments. In particular, the differential equations for the temperature ratios $\chi_i=T_i/T$ can be derived by multiplying both sides of Eq.\ \eqref{3.6} by $m_i c^2$ and integrating over velocity,
\begin{equation}
\label{3.8}
\frac{3}{2} \Lambda^* \xi^* \frac{\partial \chi_i}{\partial \xi^*}=\chi_i \Lambda^*-\Lambda_i^*,
\end{equation}
where $\Lambda^*=x_1\Lambda_1^*+x_2 \Lambda_2^*$,
\begin{equation}
\label{3.9}
\Lambda^{*}_i = 2 \omega^* \xi^{*1/3} \frac{\chi_i}{M_i^\beta} - \frac{\xi^*}{M_i^{\lambda-1}} +\chi_i \zeta_{i}^*,
\end{equation}
and $\zeta_{i}^*=\zeta_{i}/\nu_0$ is
\beq
\label{3.10}
\zeta_{i}^*=\sum_{j=1}^2 \frac{2\nu_{ij}^*}{d}\mu_{ji}(1+\alpha _{ij})\left[1-\frac{\mu_{ji}}{2}(1+\alpha_{ij})
\frac{\theta_i+\theta_j}{\theta_j}\right],
\eeq
where $\nu_{ij}^*=\nu_{ij}/\nu_0$. In the case of elastic collisions ($\zeta_i^*=0$), the steady solution ($\Lambda^*=\Lambda_1^*=\Lambda_2^*=0$) to Eq.\ \eqref{3.8} yields the result
\beq
\label{3.10.1}
T_{i,\text{el}}=\frac{\xi_\text{b}^2}{2\gamma_\text{b}m_i^{\lambda-\beta-1}}.
\eeq
Thus, $T_{1,\text{el}}=T_{2,\text{el}}$ if $m_1=m_2$ (regardless of the values of $\lambda$ and $\beta$) or $\lambda-\beta=1$ for $m_1\neq m_2$. This suggests to introduce the \emph{bath} temperature $T_\text{b}$ as
\beq
\label{3.10.1}
T_\text{b}=\frac{\xi_b^2}{2\gamma_b(2\overline m)^{\lambda-\beta-1}}.
\eeq
The bath temperature can be interpreted as the temperature of the molecular gas surrounding the solid particles. Of course, the thermostat parameters $\xi_b^2$ and $\gamma_b$ fix the value of $T_\text{b}$.

\begin{figure}
\includegraphics[width=0.5\columnwidth]{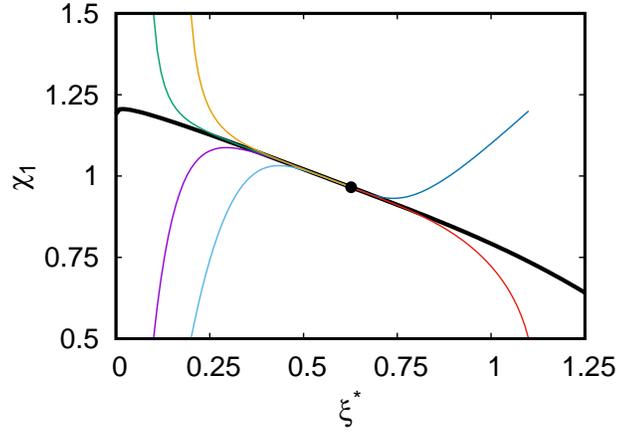}
\caption{Temperature ratio $\chi_1$ versus the (reduced) strength noise $\xi^*$ for $d=3$, $\sigma_1=\sigma_2$, $m_1/m_2=10$, $x_1=\frac{1}{2}$, and the (common) coefficient of restitution $\al_{11}=\al_{22}=\al_{12}=0.9$. Six different initial conditions are considered for $\chi_1(\xi^*)$: $\chi_1(0.1)=0.5$ (violet line), $\chi_1(0.1)=1.5$ (green line), $\chi_1(0.2)=0.5$ (light-blue line), $\chi_1(0.2)=1.5$ (yellow line), $\chi_1(1.1)=0.5$ (red line), and $\chi_1(1.1)=1.2$ (blue line). The filled circle corresponds to the value of $\chi_1$ in the steady state ($\Lambda^*=0$). Regardless of the initial condition considered we observe that the system evolves along the hydrodynamic solution (common thick black line) until the steady state is reached.}
\label{fig1}
\end{figure}
\begin{figure}
\includegraphics[width=0.5\columnwidth]{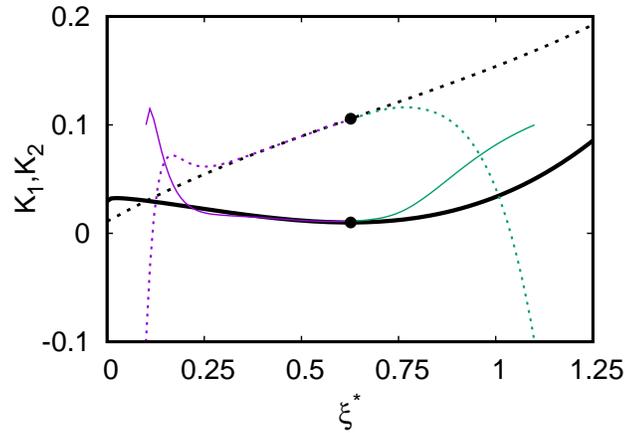}
\caption{Cumulants $K_1$ (solid lines) and $K_2$ (dotted lines) versus the (reduced) strength noise $\xi^*$ for $d=3$, $\sigma_1=\sigma_2$, $m_1/m_2=10$, $x_1=\frac{1}{2}$, and the (common) coefficient of restitution $\al_{11}=\al_{22}=\al_{12}=0.9$. Two different initial conditions are considered for $K_i(\xi^*)$: $K_1(0.2)=0.1$ (violet solid line) and $K_2(0.2)=-0.1$ (violet dotted line), and $K_1(1.1)=0.1$ (green solid line) and $K_2(1.1)=-0.1$ (green dotted line). The solid and dotted black lines correspond to the hydrodynamic values of $K_1$ and $K_2$, respectively. The filled circles correspond to the values of $K_1$ and $K_2$ in the steady state ($\Lambda^*=0$).}
\label{fig2}
\end{figure}

Since $\xi^*(t)\propto T(t)^{-3/2}$, we can take $\xi^*$ instead of the (scaled) time $t/t_0$ ($t_0$ being an arbitrary unit of time) to analyze the time-dependence of the temperature ratios. Thus, the solution to Eq.\ \eqref{3.8} provides the dependence of the temperature ratios $\chi_i(x_1,\omega^*,\xi^*)$ on the reduced noise strength $\xi^*$. Note first that Eq.\ \eqref{3.8} reduces to that of the undriven case when $\xi^*\to 0$ but keeping $\omega^*$ finite (which is equivalent to $\gamma_\text{b}\to 0$ and $\xi_\text{b}\to 0$ but keeping $\gamma_\text{b} \xi_\text{b}^{-2/3}$ finite). This physical situation could be achieved by assuming that the granular temperature is much larger than that of the bath $T_\text{b}$ and so, the dynamic of grains is not substantially affected by the presence of the bath. In this limit case ($\xi^*\to 0$), Eq.\ \eqref{3.8} reduces to
\beq
\label{3.11}
\zeta_1^*=\zeta_2^*=\zeta^*.
\eeq
Equation \eqref{3.11} is no more than the condition for determining the temperature ratio $T_1/T_2$ in the homogeneous cooling state \cite{GD99b}. The solution to Eq.\ \eqref{3.11} gives $\chi_1$ in terms of the mass and diameter ratios, the concentration $x_1$, and the coefficients of restitution. It is worthwhile to remark that the theoretical results for IMM obtained from Eq.\ \eqref{3.11} [by using the expression \eqref{3.10} for the cooling rates] exhibits an excellent agreement with those obtained from Monte Carlo simulations of IHS \cite{GA05}. On the other hand, for states close to the undriven case ($\xi^*\ll 1$), Eq.\ \eqref{3.8} admits the solution
\begin{equation}
\label{3.12}
\chi_1(x_1,\omega^*,\xi^*)\to \chi_{1,0}(x_1)+\chi_{1,1}(x_1,\omega^*) \xi^{*1/3},
\end{equation}
where $\chi_{1,0}(x_1)$ is the solution to Eq.\ \eqref{3.11} and the coefficient $\chi_{1,1}(x_1,\omega^*)$ is
\beq
\label{10}
\chi_{1,1}=-\frac{2\delta m_\beta \omega^*}{\overline{\zeta}_1^{(1)}-\overline{\zeta}_2^{(1)}+
\frac{\overline{\zeta}_1^{(0)}}{2x_2 \chi_{1,0}\chi_{2,0}}},
\eeq
with $\overline{\zeta}_1^{(0)}$ and $\overline{\zeta}_2^{(0)}$ given by Eq.\ \eqref{3.10} after the replacement $\chi_i\to \chi_{i,0}$,
\beq
\label{6}
\overline{\zeta}_1^{(1)}=\frac{\nu_{12}^*}{2d}\mu_{21}^2(1+\al_{12})^2\frac{M_1}{x_2M_2 \chi_{1,0}^2}, \quad
\overline{\zeta}_2^{(1)}=-\frac{\nu_{21}^*}{2d}\mu_{12}^2(1+\al_{12})^2\frac{M_2}{x_2M_1 \chi_{2,0}^2},
\eeq
and
\begin{equation}
\label{11}
\delta m_\beta=\frac{m_2^\beta-m_1^\beta}{(m_1+m_2)^\beta}.
\end{equation}

Beyond the above cases, we have to numerically solve Eq.\ \eqref{3.8} to get $\chi_1(x_1,\omega^*,\xi^*)$ for the homogeneous reference state. The initial condition is generated by using Eq.\ \eqref{3.12} for $\xi^*\ll 1$. This allows us to avoid the singular point $\xi^*=0$. Figure \ref{fig1} shows $\chi_1$ versus $\xi^*$ for six different initial conditions (namely, different values of $\xi^*$ and $\chi_1$). Here, as in our previous works on driven granular mixtures \cite{KG13,KG14}, $\beta=1$, $\lambda=2$, and $\omega^*\simeq 0.107$ (it corresponds to the volume fraction $0.00785$, which is of course very small). In Fig.\ \ref{fig1} it is clearly seen  that all the curves converge \emph{rapidly} towards the (common) thick black line regardless of the initial condition considered. This \emph{universal} curve is identified as the hydrodynamic solution $\chi_1(x_1,\omega^*,\xi^*)$. The steady state ($\Lambda^*=0$) is represented by the filled circle. This state was widely studied in Ref.\ \cite{KG13} where it was shown that $\chi_1$ and its derivatives are regular functions of $x_1$, $\omega^*$, and $\xi^*$. Apart from the homogeneous steady state, the transport properties in states close to the steady state were also determined in the above papers \cite{KG13,KG18,KG19}. Here, we will generalize this study by considering transport around arbitrary homogeneous time-dependent reference states, represented by the thick black line of Fig.\ \ref{fig1} in the plane $(\xi^*,\chi_1)$.

As already mentioned, even though the exact form of the distributions $\varphi_i$ is not known, their fourth cumulants (or kurtosis) $K_i$ (which measure the deviations of $\varphi_i$ from their Maxwellian form $\pi^{-d/2} \theta_i^{d/2}e^{-\theta_i c^2}$) can be exactly computed. They are defined as
\beq
\label{3.14}
K_i=2\left[\frac{4}{d(d+2)}\theta_i^2 \int \dd\mathbf{c}\; c^4 \varphi_i(\mathbf{c})-1\right].
\eeq
The evolution equation of $K_i$ can be obtained by multiplying both sides of the Boltzmann equation \eqref{3.6} by $m_i c^4$ and integrating over velocity. The calculations are long and will be omitted here for the sake of brevity. As in the case of the temperature ratio $\chi_1$, the results show that both cumulants tend to converge towards the universal hydrodynamic functions after a short transient period. This behavior is clearly illustrated in Fig.\ \ref{fig2} where $K_1(\xi^*)$ and $K_2(\xi^*)$ are plotted versus $\xi^*$  for the same initial conditions as in Fig.\ \ref{fig1}.

\section{Chapman--Enskog solution of the Boltzmann equation for IMM}
\label{sec4}

The Chapman--Enskog method \cite{CC70} generalized to inelastic collisions is applied in this section to solve the set of Boltzmann equations \eqref{2.7} for IMM up to first order in spatial gradients. The Chapman--Enskog solution will be employed then to determine the Navier--Stokes transport coefficients as functions of the coefficients of restitution, composition, the masses and diameters of grains, and the thermostat parameters.

\subsection{Sketch of the Chapman-Enskog method}

As in the conventional Chapman-Enskog method \cite{CC70}, we assume the existence of a normal solution to the Boltzmann equation in which the velocity distribution functions $f_i$ depend on space and time through a functional dependence on the hydrodynamic fields:
\beq
\label{4.1}
f_i(\mathbf{r},\mathbf{v},t)=f_i[\mathbf v|x_1(t),\mathbf U(t),T(t),p(t)].
\eeq
For small enough spatial gradients, the functional dependence \eqref{4.1} can be made explicit by expanding $f_i(\mathbf{r},\mathbf{v},t)$ in powers of a formal parameter $\epsilon$:
\begin{equation}
\label{4.2}
f_{i}=f_{i}^{(0)}+\epsilon \,f_{i}^{(1)}+\epsilon^2 \,f_{i}^{(2)}+\cdots,
\end{equation}
where each factor $\epsilon$ means an implicit gradient of the hydrodynamic fields $x_1$, $\mathbf{U}$, $p$, and $T$. The time derivatives of the fields are also expanded as $\partial_t=\partial_t^{(0)}+\epsilon\partial_t^{(1)}+\epsilon^2\partial_t^{(2)}+\cdots$. The expansion \eqref{4.2} yields similar expansions for the fluxes and the cooling rate when substituted into Eqs.\ \eqref{2.13}, \eqref{2.15}, \eqref{2.16}, \eqref{2.23}, and \eqref{2.24}:
\beq
\label{4.3.1}
\mathbf{j}_i=\mathbf{j}_i^{(0)}+\epsilon\, \mathbf{j}_i^{(1)}+\cdots, \quad \mathsf{P}=\mathsf{P}^{(0)}+\epsilon\, \mathsf{P}^{(1)}+\cdots,
\quad \mathbf{q}=\mathbf{q}^{(0)}+\epsilon\, \mathbf{q}^{(1)}+\cdots, \quad \zeta=\zeta^{(0)}+\epsilon\, \zeta^{(1)}+\cdots.
\eeq
In addition, since the partial temperatures $T_i$ are not hydrodynamic quantities, they must be also expanded in powers of the gradients as \cite{KG19,GGG19b}
\beq
\label{4.3.3}
T_i=T_i^{(0)}+\epsilon\,T_i^{(1)}+\cdots
\eeq
On the one hand, the action of the time derivatives $\partial_t^{(k)}$ on $x_1$, $\mathbf{U}$, $p$, and $T$ can be obtained from the balance equations \eqref{2.19}--\eqref{2.21} after taking into account the expansions \eqref{4.3.1}--\eqref{4.3.3}. With respect to the thermostat parameters $\gamma_\text{b}$ and $\xi_\text{b}^2$ and the difference $\Delta \mathbf{U}$, as in our previous study on IHS \cite{KG13} we assume that $\gamma_\text{b}$ and $\xi_\text{b}^2$ are of zeroth order in the gradients while $\Delta \mathbf{U}$ is considered at least to be of first order in the gradients. More details on the ordering of the terms in the kinetic equations can be found in Ref.\ \cite{KG13}.

As usual \cite{CC70}, the hydrodynamic fields $x_1$, $p$, $T$, and $\mathbf U$ are defined by the zeroth-order distributions, hence
\beq
\label{4.3.4}
\int \dd\mathbf{v}\left(f_i-f_i^{(0)}\right)=0,
\eeq
\beq
\label{4.3.5}
\sum_{i=1}^2\int \dd\mathbf{v}\; \left\{m_i\mathbf{v}, \frac{m_i}{2}V^2\right\}\left(f_i-f_i^{(0)}\right)=\left\{\mathbf{0},0\right\}.
\eeq
Since the constraints \eqref{4.3.4} and \eqref{4.3.5} must hold at any order in $\epsilon$, the remainder of the expansion must obey the orthogonality conditions
\beq
\label{4.3.6}
\int \dd\mathbf{v}f_i^{(k)}=0,
\eeq
and
\beq
\label{4.3.7}
\sum_{i=1}^2\int \dd\mathbf{v}\; \left\{m_i\mathbf{v}, \frac{m_i}{2}V^2\right\}f_i^{(k)}=\left\{\mathbf{0},0\right\},
\eeq
for $k\geq 1$. A consequence of Eq.\ \eqref{4.3.6} is that the partial densities are of zeroth order while Eq.\ \eqref{4.3.7} yields the relations
\begin{equation}
\label{eq:pr1}
\mathbf{j}_1^{(k)}=-\mathbf{j}_2^{(k)}, \quad n_1 T_1^{(k)}=-n_2 T_2^{(k)},
\end{equation}
for $k\geq 1$.

\subsection{Zeroth-order solution}

In the zeroth order, $f_i^{(0)}$ obeys the Boltzmann equation \eqref{3.1} with the replacements $\mathbf{v}\to \mathbf{V}(\mathbf{r},t)=\mathbf{v}-\mathbf{U}(\mathbf{r},t)$,  $\partial_t\to \partial_t^{(0)}$, $x_1\to x_1(\mathbf r,t)$, $T\to T(\mathbf r,t)$, and $p\to p(\mathbf r,t)$. The balance equations to zeroth order are
\beq
\label{4.4}
\partial_t^{(0)}x_1=\partial_t^{(0)}U_i=0, \quad T^{-1}\partial_t^{(0)}T=p^{-1}\partial_t^{(0)}p=-\Lambda^{(0)},
\eeq
where
\beq
\label{4.5}
\Lambda^{(0)} =  \zeta^{(0)}+2\gamma_\text{b}\sum_{i=1}^2 \frac{x_i \chi_i}{ m_i^\beta}-\frac{\xi_\text{b}^2}{p}\sum_{i=1}^2\frac{\rho_i}{m_i^\lambda}.
\eeq
Here, $\chi_i=T_i^{(0)}/T$ and $\zeta^{(0)}=\sum_i\; x_i \chi_i \zeta_i^{(0)}$ where
\beq
\label{4.6}
\zeta_i^{(0)}=\frac{2\nu_{ij}}{d}\mu_{ji}(1+\alpha_{ij})\left[1-\frac{\mu_{ji}}{2}(1+\alpha_{ij})
\frac{\theta_i+\theta_j}{\theta_j}\right].
\eeq
We recall that $\nu_{ij}$ is defined by Eq.\ \eqref{2.17} with $T_i=T_i^{(0)}$. The velocity distribution $f_i^{(0)}$ is given by the scaling \eqref{3.4} except that now the hydrodynamic fields are local quantities. Since $f_i^{(0)}$ is isotropic in $\mathbf{V}$, it follows that
\beq
\label{4.6.1}
\mathbf{j}_i^{(0)}=\mathbf{q}^{(0)}=\mathbf{0}, \quad P_{k\ell}^{(0)}=p\delta_{k\ell},
\eeq
where $p=nT$.

\section{First-order solution. Navier--Stokes transport coefficients}
\label{sec5}

The first-order contributions to the distribution functions are considered in this section. Since the mathematical steps involved in the determination of $f_i^{(1)}(\mathbf{V})$ are quite similar to those made in Ref.\ \cite{KG13} for IHS, the derivation is omitted, and we refer the interested reader to the Appendix B of \cite{KG13} for specific details. The only subtle point not accounted for in Ref.\ \cite{KG13}, but recognized later in an erratum \cite{KG19}, is the existence of non vanishing contributions to the partial temperatures $T_i^{(1)}$.

Taking into account the contributions to $f_i^{(1)}(\mathbf{V})$ coming from $T_i^{(1)}$, the kinetic equation of $f_i^{(1)}$ can be written as
\beqa
\label{4.14}
\nonumber
& & \partial_t^{(0)}f_{1}^{(1)}-\frac{\gamma_\text{b}}{m_1^\beta} \frac{\partial}{\partial\mathbf{ v}}\cdot \mathbf{V} f_1^{(1)}-\frac{1}{2}\frac{\xi_\text{b}^2}{m_1^\lambda}\frac{\partial^2 f_1^{(1)}}{\partial v^2} +{\mathcal L}_{1} f_{1}^{(1)} +{\mathcal M}_{1}f_{2}^{(1)}=\mathbf{A}_{1}\cdot \nabla x_{1}+\mathbf{ B}_{1}\cdot \nabla p+\mathbf{ C}_{1}\cdot \nabla T\\ & & +D_{1,k\ell}\frac{1}{2}\left(\nabla_{k}U_{\ell}+\nabla_{\ell}U_{k}-\frac{2}{d}\delta_{k\ell}\nabla \cdot\mathbf{U}\right)+E_1 \nabla \cdot \mathbf{U}+\mathbf{G}_1\cdot \Delta \mathbf{U},
\eeqa
where
\beq
\label{4.15}
\mathbf{ A}_{1}(\mathbf{V})=-\mathbf{V}\frac{\partial f_1^{(0)}}{\partial x_1}+\frac{\gamma_\text{b}  m_1 m_2\delta m_\beta}{\rho^2 \overline m^\beta}\frac{p}{T} D \frac{\partial f_1^{(0)}}{\partial \mathbf{V}},  \quad
\mathbf{ B}_{1}(\mathbf{V})=-\mathbf{V}\frac{\partial f_1^{(0)}}{\partial p} -\rho^{-1} \frac{\partial f_1^{(0)}}{\partial \mathbf{V}} +\frac{\gamma_\text{b}\delta m_\beta}{p\overline m^\beta} D_p \frac{\partial f_1^{(0)}}{\partial \mathbf{V}},
\eeq
\beq
\label{4.16}
\mathbf{C}_{1}(\mathbf{V})=-\mathbf{V} \frac{\partial f_1^{(0)}}{\partial T}+\frac{\gamma_\text{b}\delta m_\beta}{T\overline m^\beta}D_T \frac{\partial f_1^{(0)}}{\partial \mathbf{V}},  \quad D_{1,k\ell}(\mathbf{V})=V_k \frac{\partial f_1^{(0)} }{\partial V_\ell},
\eeq
\beq
\label{4.17}
E_1(\mathbf{V})=\left(\frac{2}{d}+\zeta_U+2\gamma_bx_1\chi_U\frac{\delta m_\beta}{\overline m^\beta}\right)\left(p\partial_p+T\partial_T\right)f_1^{(0)}+p \frac{\partial f_1^{(0)}}{\partial p}+\frac{1}{d}\mathbf{V}\cdot \frac{\partial f_1^{(0)}}{\partial \mathbf{V}},
\eeq
\beq
\label{4.18}
\mathbf{G}_1(\mathbf{V})=\frac{\gamma_\text{b}}{\rho} \frac{\delta m_\beta}{\overline m^\beta}
\left(\rho_2+ D_U\right) \frac{\partial f_1^{(0)}}{\partial \mathbf{V}}.
\eeq
In Eq.\ \eqref{4.14}, the linear operators ${\mathcal L}_{1}$ and ${\mathcal M}_{1}$ are defined as
\beq
\label{4.8}
{\mathcal L}_1X=-J_{11}[\mathbf v|f_1^{(0)},X]-J_{11}[\mathbf v|X,f_1^{(0)}]-J_{12}[\mathbf v|X,f_2^{(0)}],
\quad {\mathcal M}_1X=-J_{12}[\mathbf v|f_1^{(0)},X].
\eeq
where $X(\mathbf v$) is a generic function of the velocity. The kinetic equation for $f_2^{(1)}$ can be easily obtained from Eq.\ \eqref{4.14} by just making the changes $1\leftrightarrow 2$. Upon writing Eqs.\ \eqref{4.15}--\eqref{4.16} use has been made of the constitutive equation for the mass flux $\mathbf{j}_1^{(1)}$. It is given by \cite{KG13}
\begin{equation}
\label{4.13}
\mathbf{j}_{1}^{(1)}=-\frac{m_{1}m_{2}p}{\rho T } D\nabla x_{1}-\frac{\rho}{p}D_{p}\nabla p-\frac{\rho}{T}D_{T}\nabla T-D_U \Delta \mathbf{U},
\end{equation}
where $D$ is the diffusion coefficient, $D_p$ is the pressure diffusion coefficient, $D_{T}$ is the thermal diffusion coefficient, and $D_U$ is the velocity diffusion coefficient. In addition, to get the expression \eqref{4.17} for $E_1(\mathbf{V})$, we have taken into account that the scalar quantities
$T_i^{(1)}$ and $\zeta^{(1)}$ can only be coupled to the divergence of the flow velocity field $\nabla \cdot \mathbf{U}$. As a consequence,
\beq
\label{4.12}
T_1^{(1)}=\frac{T}{\nu_0} \chi_U \nabla \cdot \mathbf{U}, \quad \zeta^{(1)}=\zeta_U \nabla \cdot \mathbf{U},
\eeq
where $\chi_U$ and $\zeta_U$ are dimensionless quantities to be determined. Since $n_1 T_1^{(1)}=-n_2 T_2^{(1)}$, then $T_2^{(1)}=-(x_1 T/x_2 \nu_0)\chi_U \nabla \cdot \mathbf{U}$, where $\nu_0$ is the effective collision frequency defined in Eq.\ \eqref{2.18}.

It is worth noting that Eqs.\ \eqref{4.14}--\eqref{4.18} are similar to those obtained for IHS \cite{KG13,KG19}, except for the explicit forms of the terms $E_i(\mathbf{V})$ and the linearized Boltzmann collision operators $\mathcal{L}_i$ and $\mathcal{M}_i$. However, the road map for determining the transport coefficients for IMM is different to that for IHS. An important advantage of using the forms of $\mathcal{L}_i$ and $\mathcal{M}_i$ of IMM is that the Navier--Stokes transport coefficients can be \emph{exactly} obtained from the Boltzmann collisional moments associated with $m_i \mathbf{V}$, $m_i \mathbf{V}\mathbf{V}$, and $\frac{m_i}{2} V^2 \mathbf{V}$ \cite{GA05}. This contrasts with the results derived for IHS \cite{KG13} where the transport coefficients were approximately determined by truncating a series expansion of the distribution functions $f_i^{(1)}(\mathbf{V})$ in Sonine polynomials.

Once the kinetic equations verifying the distributions $f_i^{(1)}$ are known, the set of Navier--Stokes transport coefficients of the granular binary mixture can be obtained. While the mass flux $\mathbf{j}_1^{(1)}$ is defined by Eq.\ \eqref{4.13}, the pressure tensor is
\beq
\label{5.1}
P_{k\ell}^{(1)}= -\eta\left(\frac{\partial U_i}{\partial r_j}+\frac{\partial U_j}{\partial r_i}-\frac{2}{d}\delta_{k\ell}\nabla \cdot \mathbf{U}\right),
\eeq
and the heat flux is
\beq
\label{5.2}
\mathbf q^{(1)}=-T^2 D^{\prime \prime}\nabla x_1-L\nabla p-\kappa\nabla T-\kappa_{U}\Delta\mathbf U.
\eeq
In Eqs.\ \eqref{5.1} and \eqref{5.2}, $\eta$ is the shear viscosity coefficient, $D^{\prime \prime}$ is the Dufour coefficient, $L$ is the pressure energy coefficient, $\kappa$ is the thermal conductivity, and $\kappa_U$ is the velocity conductivity.

The evaluation of the transport coefficients as well as the first-order contribution to the partial temperatures follows similar mathematical steps to those made in the case of IHS \cite{KG13}. Since these calculations are standard in organization (although somewhat complex in execution), we relegate the long and tedious technical details of these calculations to the Appendices \ref{appA} and \ref{appB}. As expected, the time-dependent forms of the set of transport coefficients $\left\{D, D_p, D_T, D_U, \eta, D'', L, \kappa, \kappa_U\right\}$ are given in terms of the solutions of nonlinear differential equations in the (reduced) variable $\xi^*$. Only simple analytical solutions to these equations are obtained in the cases of undriven granular mixtures ($\xi^*=0$) and driven granular mixtures under steady state conditions ($\Lambda^{(0)}=0$). These two cases will be separately considered to perform a comparison with previous results obtained for IHS \cite{GD02,GA05,KG13}.

\section{Time-dependent transport coefficients. Comparison with IHS}
\label{sec6}

\subsection{Unsteady hydrodynamic solution}

Although most of the works devoted on transport in driven granular gases have been focused in the steady state, it is also worthwhile  studying the time-dependent forms of the transport coefficients. As said in the Introduction, this is in fact one of the new added values of the present contribution. As we discussed in section \ref{sec3}, after a transient kinetic regime, we expect that the mixture achieves an \emph{unsteady} hydrodynamic state \cite{K18,GMT12} where the (scaled) transport coefficients ($D^*$, $D_p^*$, $D_T^*$, $\ldots$) depend on time only through the reduced parameter $\xi^*$. The definitions of the above \emph{scaled} transport coefficients are given in Eqs.\ \eqref{5.7} and \eqref{b2} of the Appendix \ref{appA}. In the sequel, we illustrate the $\xi^*$--dependence of the (scaled) transport coefficients for different values of the parameter space of the system.

\begin{figure}[!h]
\centering
\includegraphics[width=0.45\textwidth]{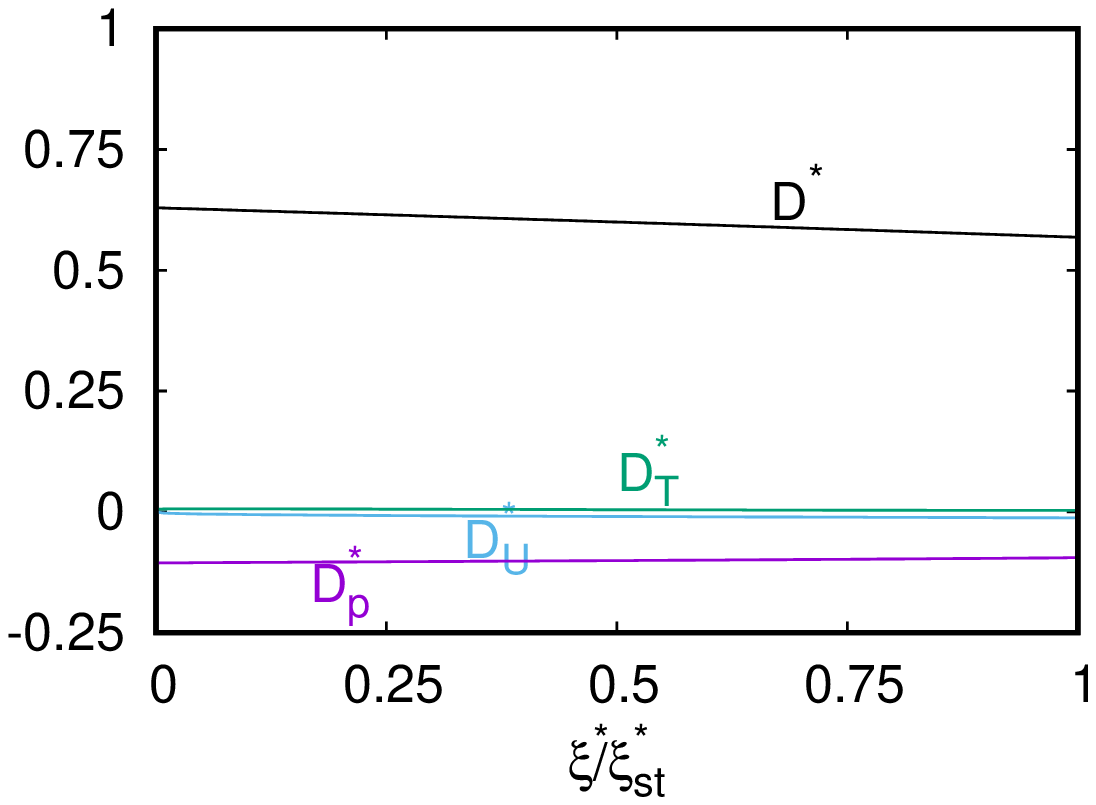}
\includegraphics[width=0.45\textwidth]{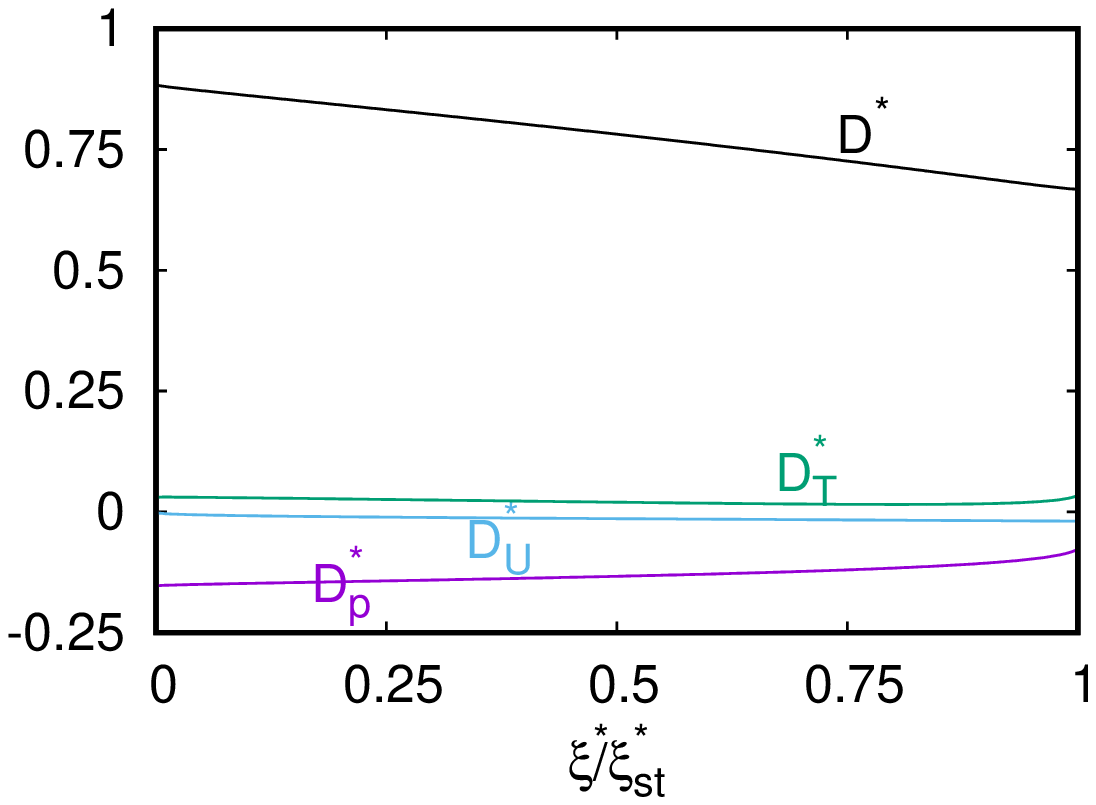}
\caption{Plot of the dimensionless diffusion transport coefficients as a function of $\xi^*/\xi_{\text{st}}^*$ for $d=2$, $\sigma_1=\sigma_2$, $m_1/m_2=2$, $x_1=\frac{1}{2}$, and two different values of the (common) coefficient of restitution:  $\al=0.9$ (left panel) and $\al=0.6$ (right panel). The black, green, blue, and violet lines correspond to the coefficients $D^*$, $D_{T}^*$, $D_U^*$, and $D_p^*$, respectively. Note that the value of $\xi_{\text{st}}^*$ is different in each panel.}
\label{fig3}
\end{figure}
\begin{figure}[!h]
\centering
\includegraphics[width=0.45\textwidth]{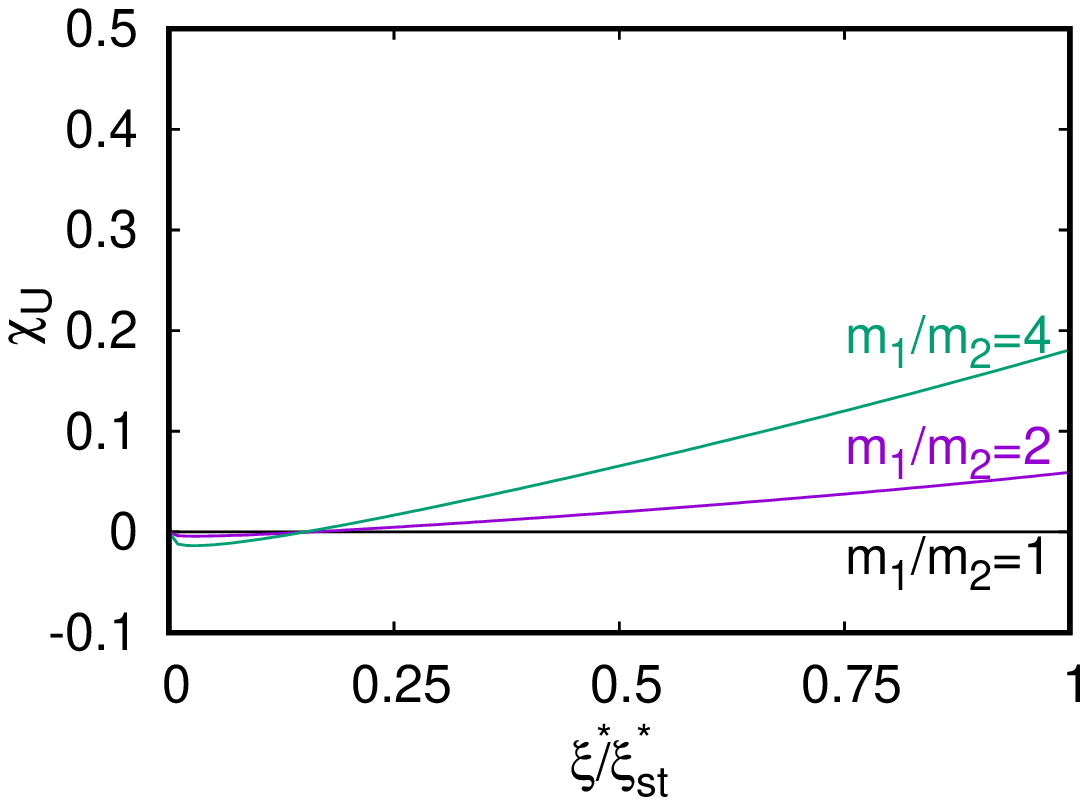}
\includegraphics[width=0.45\textwidth]{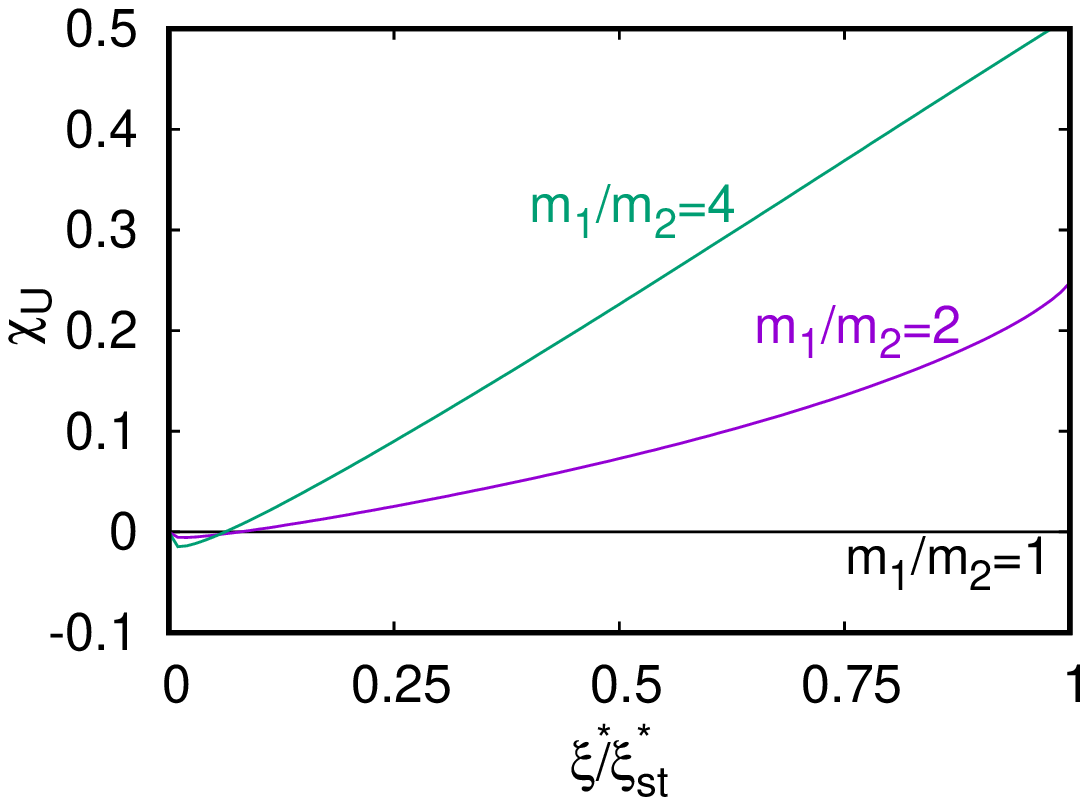}
\caption{Plot of the coefficient $\chi_U$ of the first-order contribution to the partial temperatures as a function of $\xi^*/\xi_{\text{st}}^*$ for $d=2$, $\sigma_1=\sigma_2$, $x_1=\frac{1}{2}$, and three different values of the mass ratio $m_1/m_2$. Two different values of the (common) coefficient of restitution are considered:  $\al=0.9$ (left panel) and $\al=0.6$ (right panel). Note that the value of $\xi_{\text{st}}^*$ is different in each panel.}
\label{fig9}
\end{figure}
\begin{figure}[!h]
\centering
\includegraphics[width=0.45\textwidth]{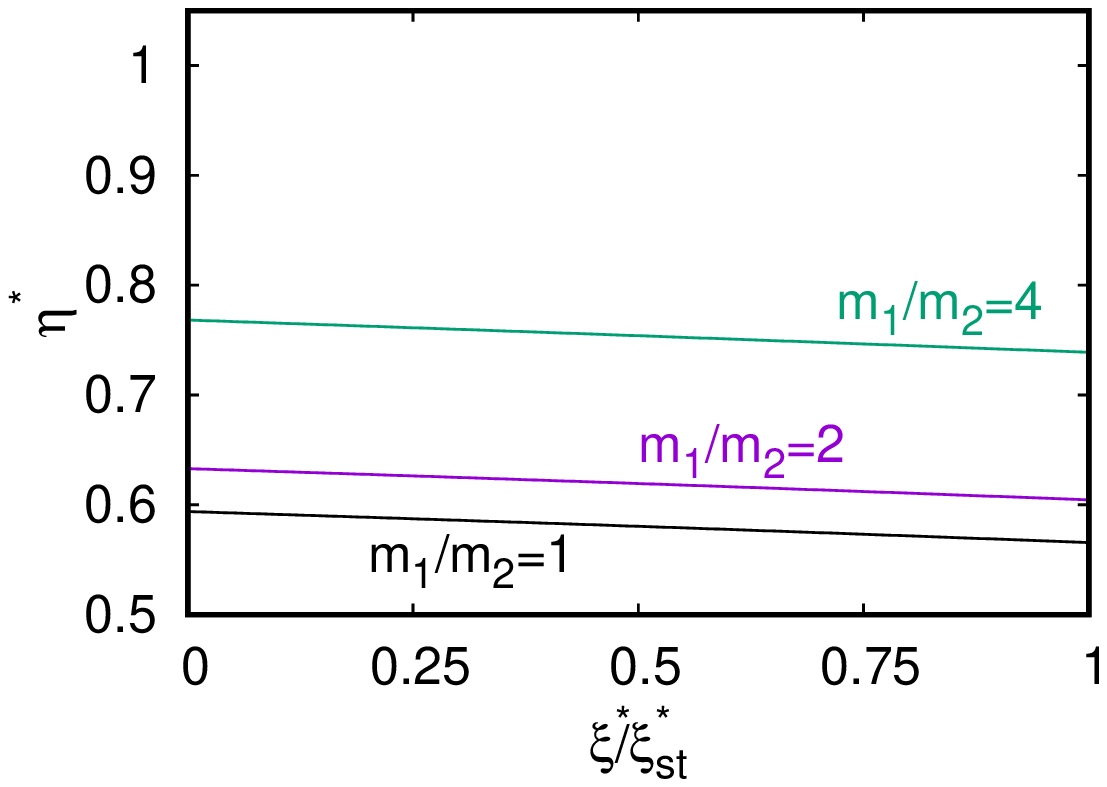}
\includegraphics[width=0.45\textwidth]{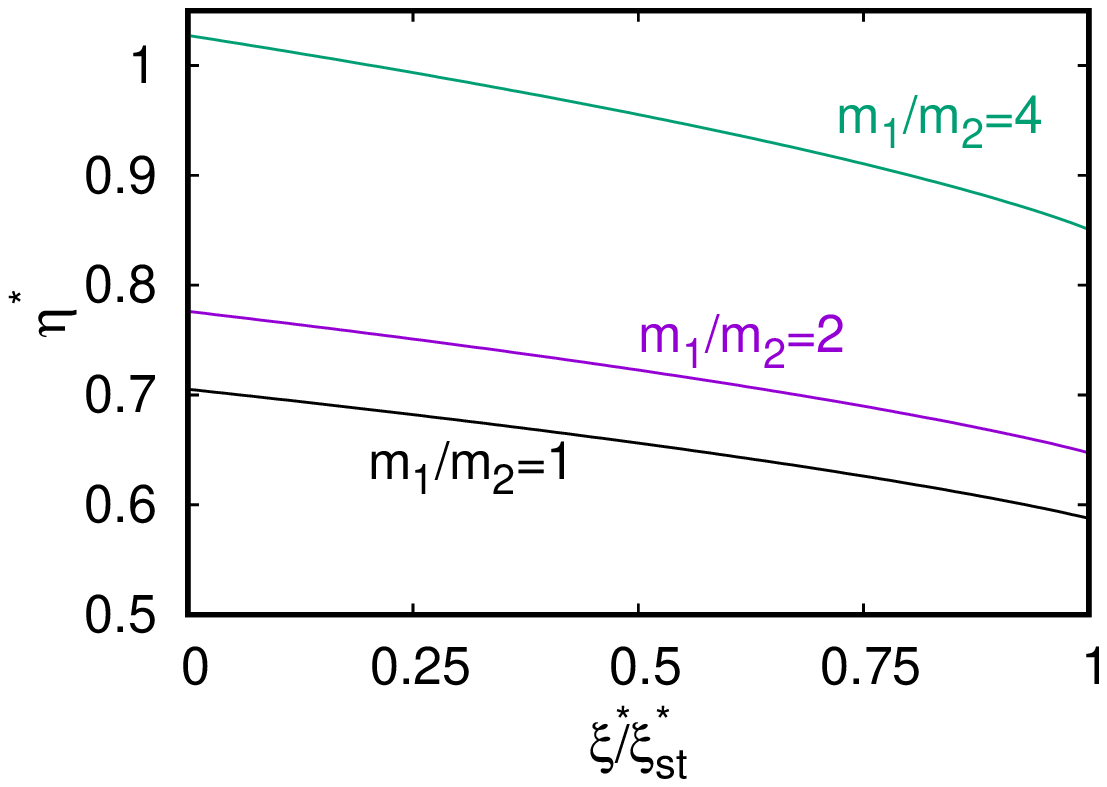}
\caption{Plot of the dimensionless shear viscosity $\eta^*$ as a function of $\xi^*/\xi_{\text{st}}^*$ for $d=2$, $\sigma_1=\sigma_2$, $x_1=\frac{1}{2}$, and three different values of the mass ratio $m_1/m_2$. Two different values of the (common) coefficient of restitution are considered:  $\al=0.9$ (left panel) and $\al=0.6$ (right panel). Note that the value of $\xi_{\text{st}}^*$ is different in each panel.}
\label{fig4}
\end{figure}

As in our previous papers \cite{KG13,KG18} on driven granular mixtures, we are here mainly interested in studying the dependence of the (scaled) transport coefficients on inelasticity. To capture this dependence, the above coefficients are normalized with respect to their values for elastic collisions. In addition, only the simplest case of a common coefficient of restitution ($\al_{11}=\al_{22}=\al_{12}\equiv \al$) of an equimolar binary mixture ($x_1=\frac{1}{2}$) with the same diameters ($\sigma_1=\sigma_2$) and with parameters $\beta=1$, $\lambda=2$ is considered. In addition, we take a volume fraction of $0.00785$, which corresponds to a very dilute system. The value of $\omega^*$ for this system can be easily inferred from table \ref{tab:1}. The above choice of parameters reduces the parameter set to three quantities: $\left\{\xi^*, m_1/m_2, \al \right\}$.

%\subsection{Influence of the thermostat}

%\end{document}
Figure \ref{fig3} shows the dependence of the scaled diffusion coefficients ($D^*$, $D_p^*$, $D_T^*$, and $D_U^*$; they are defined in Eq.\ \eqref{5.7}) on the scaled parameter $\xi^*/\xi_{\text{st}}^*$ for a two-dimensional granular mixture with $m_1/m_2=2$ and two different values of the coefficient of restitution: $\al=0.9$ (left panel) and $\al=0.6$ (right panel). The steady value $\xi_{\text{st}}^*$ is obtained from the condition $x_1\Lambda_1^*+x_2\Lambda_2^*=0$. It corresponds to the value of $\xi^*$ where the density and granular temperature reach the local values imposed by the thermostat. Note that we restrict our study on the unsteady solution to the interval between the undriven state ($\xi^*=0$) and the asymptotic final steady state ($\xi^*/\xi_{\text{st}}^*=1)$. The undriven state can be achieved either because both parameters $\gamma_b$ and $\xi_b^2$ go to zero (keeping $\gamma_\text{b} \xi_\text{b}^{-2/3}$ finite) or because the granular temperature is big enough ($T\gg T_\text{b}$). As expected, Fig.\ \ref{fig3} shows that the influence of the thermostat (as measured by the difference between the values of the dimensionless diffusion coefficients with and without a thermostat) is more significant as the inelasticity increases. This is quite apparent in the right panel of Fig.\ \ref{fig3}, specially in the case of the diffusion coefficient $D^*$.

The coefficient $\chi_U$ is plotted in Fig.\ \ref{fig9} as a function of $\xi^*/\xi_{\text{st}}^*$ for $\al=0.9$ and 0.6 and different values of the mass ratio $m_1/m_2$. This coefficient is defined by Eq.\ \eqref{4.12} and provides the first-order contribution to the partial temperature $T_1$. Although this coefficient was calculated many years ago \cite{KS79b} for dense molecular mixtures and more recently, for dense granular mixtures \cite{GGG19b,GGKG20}, we do not aware of any previous calculation of $\chi_U$ for low-density driven granular mixtures.
As expected, $\chi_U$ vanishes (i) for $\xi^*=0$ (undriven case) \cite{GD02} and (ii) for mechanically equivalent particles ($\sigma_1=\sigma_2$, $m_1=m_2$ and $\al_{ij}=\al$). We observe that $\chi_U$ is negative near $\xi^*=0$ and then it becomes positive for larger values of $\xi^*$. It is also quite apparent that $\chi_U$ exhibits a non-monotonic dependence on $\xi^*$ since it decreases (increases) with increasing $\xi^*$ for $\xi^* \lesssim 0.05$ ($\xi^* \gtrsim 0.05$). In addition, for strong inelasticity, we observe that the influence of $\xi^*$ on $\chi_U$ increases with the mass ratio.  The dependence of the (reduced) shear viscosity $\eta^*=(\nu_0/p)\eta$ on $\xi^*/\xi_{\text{st}}^*$ is plotted in Fig.\ \ref{fig4}. We infer similar conclusions to those found before for the diffusion transport coefficients. On the one hand, at a given value of the coefficient of restitution, the impact of the (scaled) parameter $\xi^*$ on $\eta^*$ is more noticeable as the mass ratio increases. On the other hand, at a given value of the mass ratio, the bigger the inelasticity the more the influence of the thermostat is.
Similar conclusions are obtained for the (scaled) heat flux transport coefficients.

\subsection{Comparison with the transport coefficients of IHS: Undriven and driven steady solutions}

Apart from analyzing the dependence of transport coefficients on $\xi^*$, the exact analytical results derived here in the undriven and driven steady states allows us to gauge the degree of reliability of IMM via a comparison with previous results derived for IHS in both situations by considering the so-called leading Sonine approximation. To the best of our knowledge, the only comparison between IMM and IHS for granular mixtures has been performed in Ref.\ \cite{GA05} for the diffusion coefficients in the case of undriven mixtures and in Refs.\ \cite{G03bis,GT15} for non-Newtonian transport coefficients. Here, we extend such comparison between both interaction models by considering the complete set of Navier--Stokes transport coefficients for $\xi^*=0$ (free cooling mixtures) and $\xi^*=\xi_{\text{st}}^*$ (driven mixtures under steady conditions).

\begin{figure}[!h]
\centering
\includegraphics[width=0.45\textwidth]{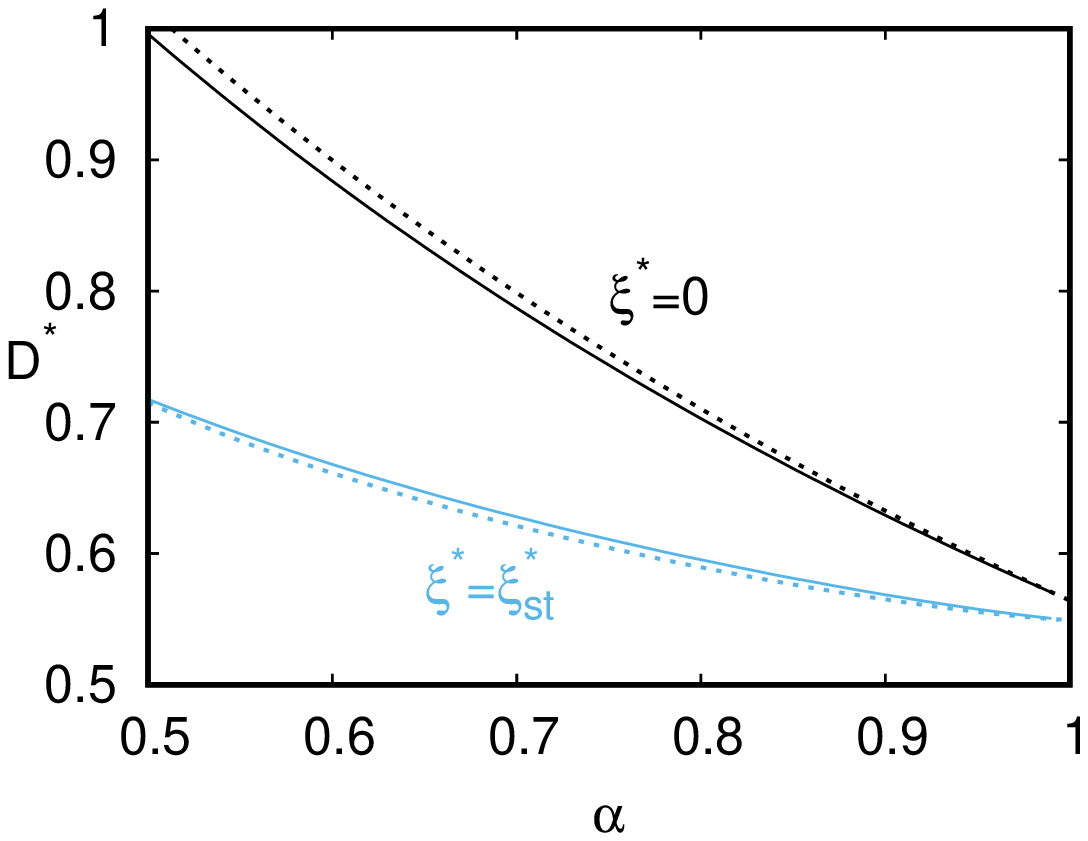}
\includegraphics[width=0.45\textwidth]{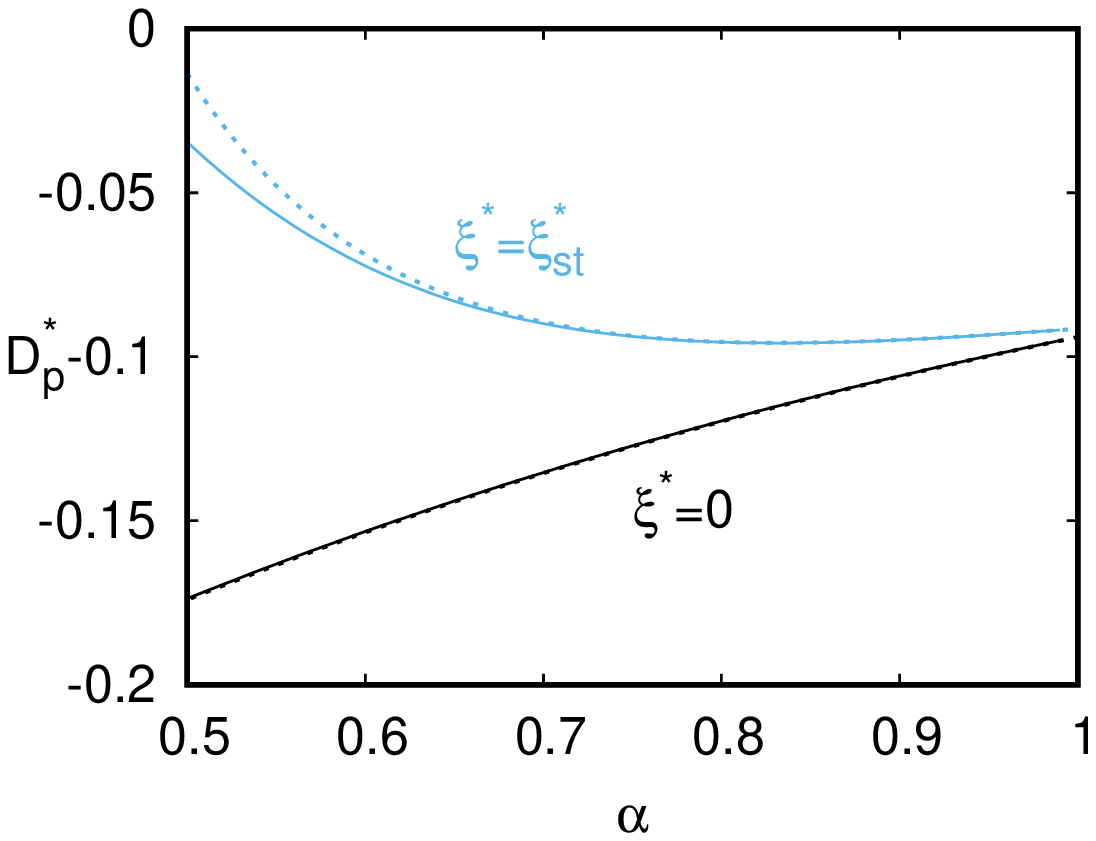}
\includegraphics[width=0.45\textwidth]{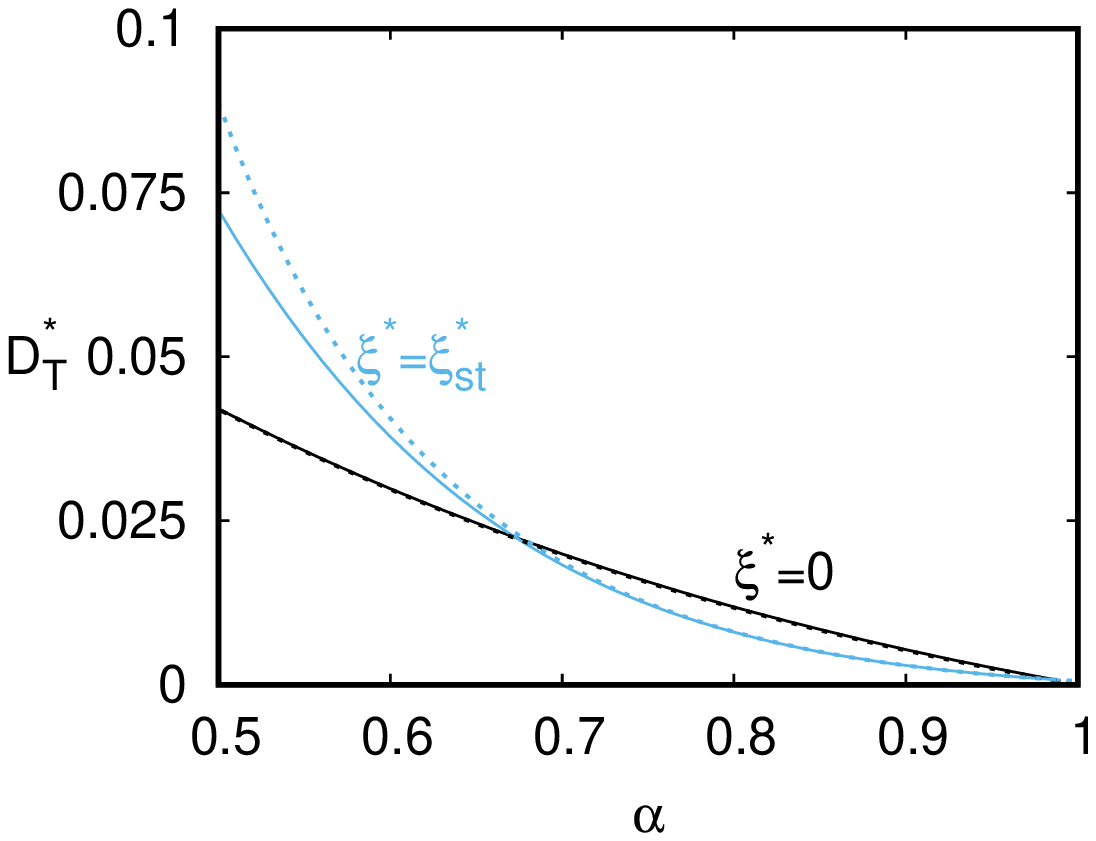}
\includegraphics[width=0.45\textwidth]{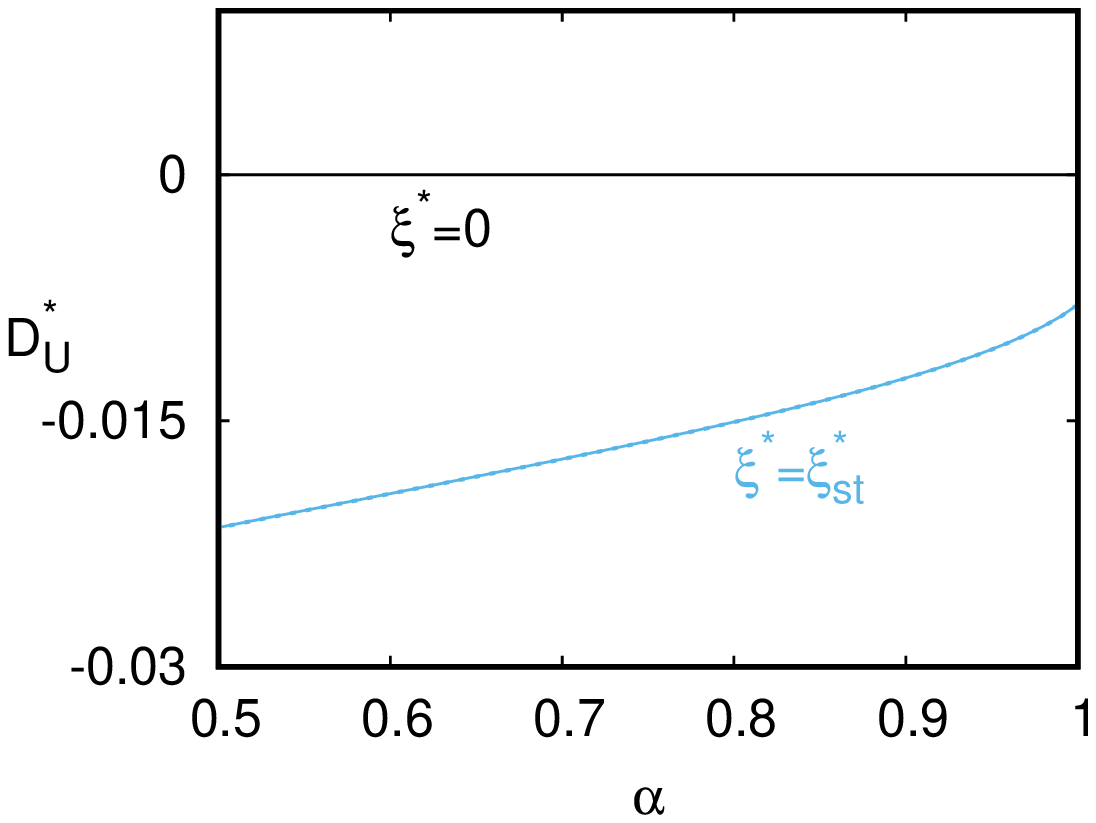}
\caption{Plot of the dimensionless diffusion transport coefficients versus the (common) coefficient of restitution $\alpha$ for hard disks ($d=2$) with $x_1=\frac{1}{2}$, $m_1/m_2=2$, and $\sigma_1/\sigma_2=1$. We consider both driven ($\xi^*=\xi_{\text{st}}^*$) and undriven ($\xi^*=0$) granular mixtures. The solid lines refer to the results derived here for IMM while the dotted lines correspond to the results obtained for IHS in Ref.\ \cite{KG13} in the first-Sonine approximation.}
\label{fig5}
\end{figure}

The dimensionless diffusion transport coefficients are plotted in Fig.\ \ref{fig5} versus $\al$ for hard disks ($d=2$) with $x_1=\frac{1}{2}$, $m_1/m_2=2$, and $\sigma_1/\sigma_2=1$. We include the results obtained for IHS (dotted lines) \cite{GD02,KG13}. Figure \ref{fig5} highlights the excellent agreement found between the predictions of the first Sonine approximation for IHS and the exact results for IMM in the whole range of values of $\al$ analyzed. We have seen that this excellent agreement is kept when we consider other type of systems (disparate masses and/or strong inelasticity).

\begin{figure}[!h]
\centering
\includegraphics[width=0.45\textwidth]{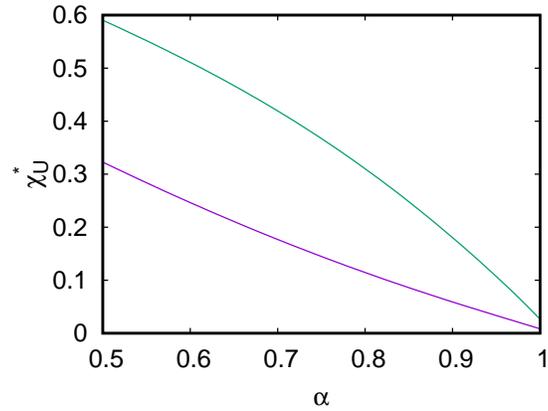}
\caption{Plot of the coefficient $\chi_U$ of the first-order contribution to the partial temperatures versus the (common) coefficient of restitution $\alpha$  for $d=2$, $\sigma_1=\sigma_2$, $x_1=\frac{1}{2}$, and $m_1/m_2=2$ (violet line) and $m_1/m_2=4$ (green line). The conditions of the steady state ($\xi^*=\xi^*_{\text{st}}$) are considered here.}
\label{fig10}
\end{figure}
\begin{figure}[!h]
\centering
\includegraphics[width=0.45\textwidth]{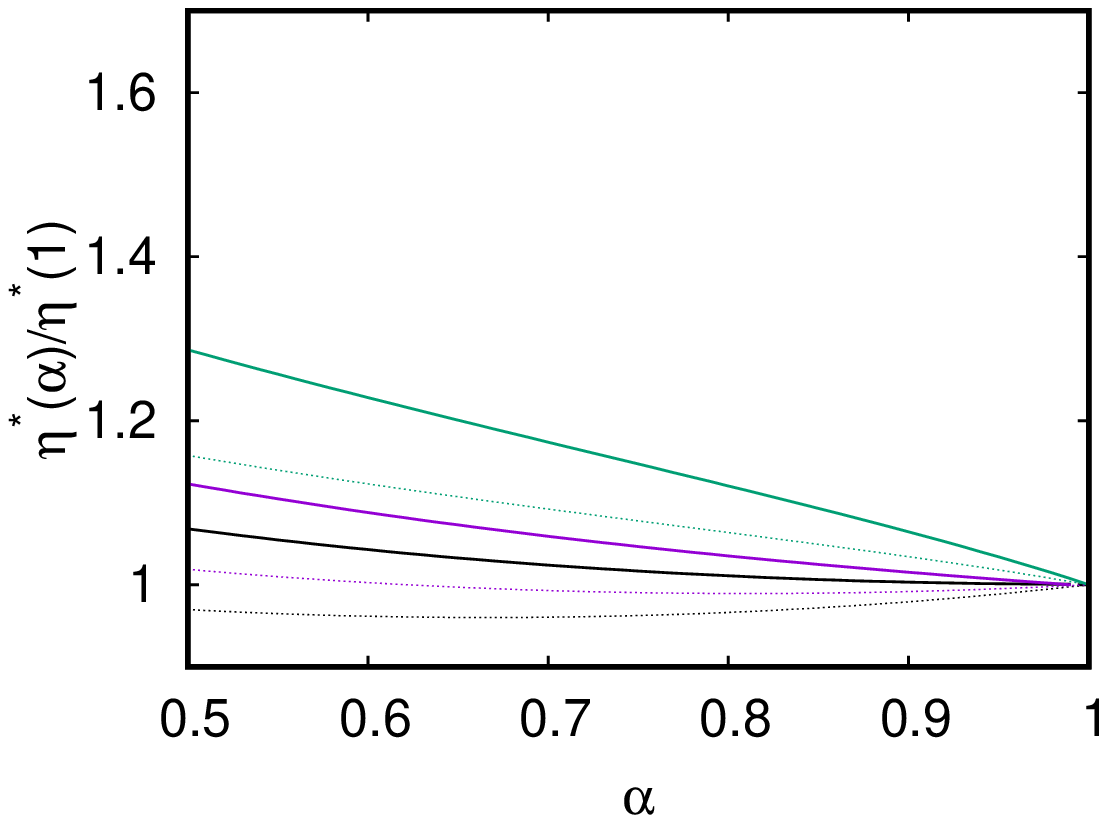}
\includegraphics[width=0.45\textwidth]{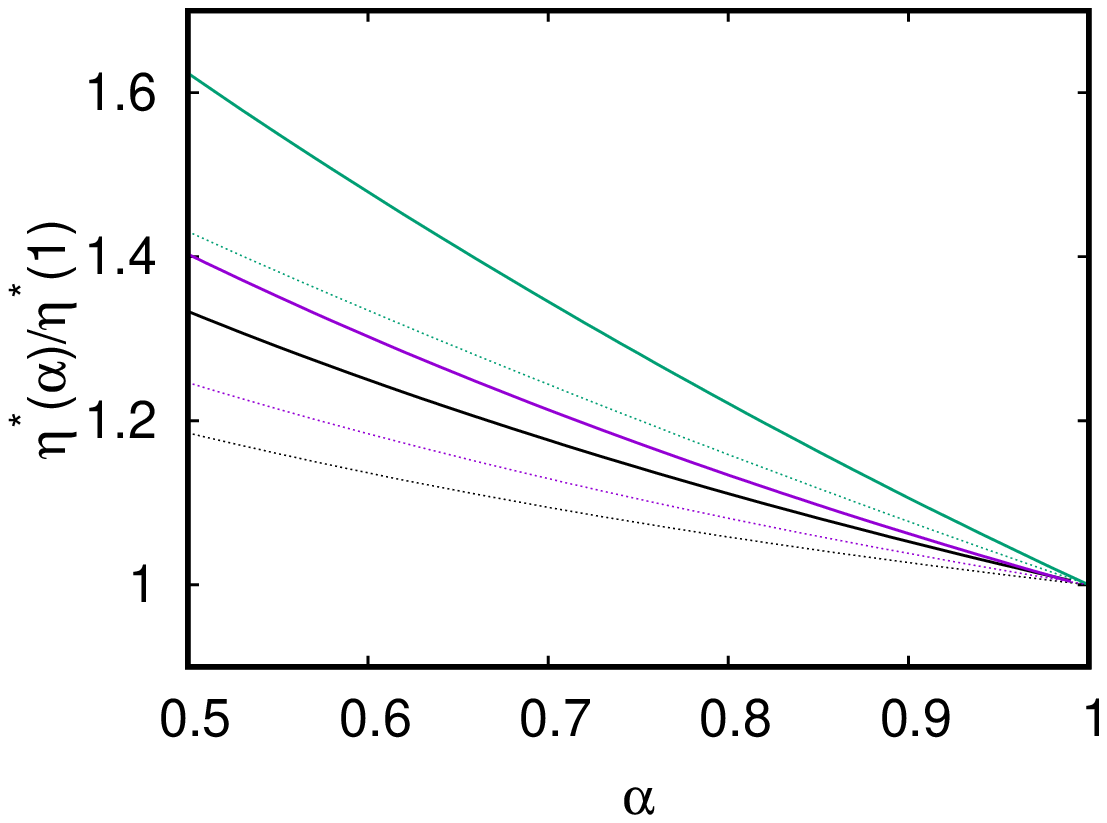}
\caption{Plot of the scaled shear viscosity coefficient $\eta^*(\al)/\eta^*(1)$ as a function of the (common) coefficient of restitution $\alpha$ for hard disks ($d=2$) with $x_1=\frac{1}{2}$ and $\sigma_1/\sigma_2=1$. Three different values of the mass ratio are considered: $m_1/m_2=1$ (black lines), $m_1/m_2=2$ (violet lines), and $m_1/m_2=4$ (green lines). We consider both driven granular mixtures under steady conditions ($\xi^*=\xi_{\text{st}}^*$, left panel) and undriven granular mixtures ($\xi^*=0$, right panel). The solid lines refer to the results derived in this paper for IMM while the dotted lines correspond to the results obtained for IHS in Ref.\ \cite{KG13} in the first-Sonine approximation. Here, $\eta^*(1)$ is the value of the shear viscosity when the collision are elastic.}
\label{fig6}
\end{figure}
\begin{figure}[!h]
\centering
\includegraphics[width=0.45\textwidth]{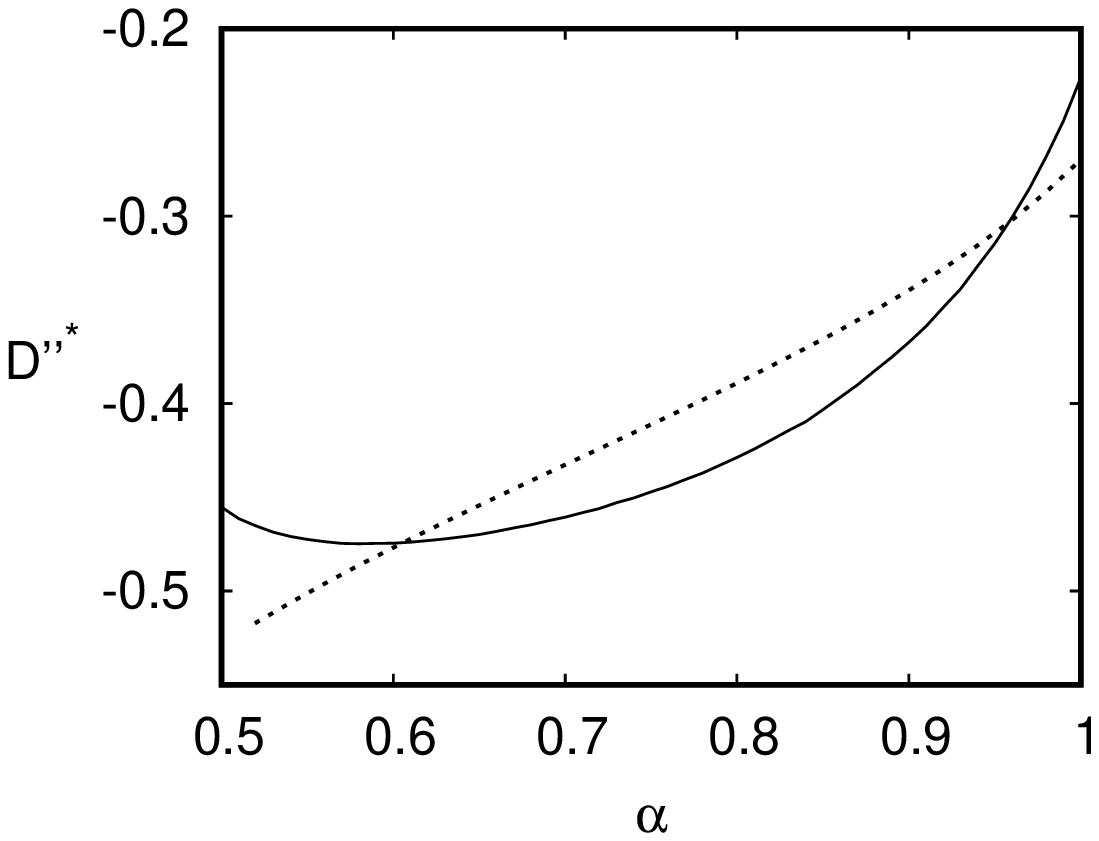}
\includegraphics[width=0.45\textwidth]{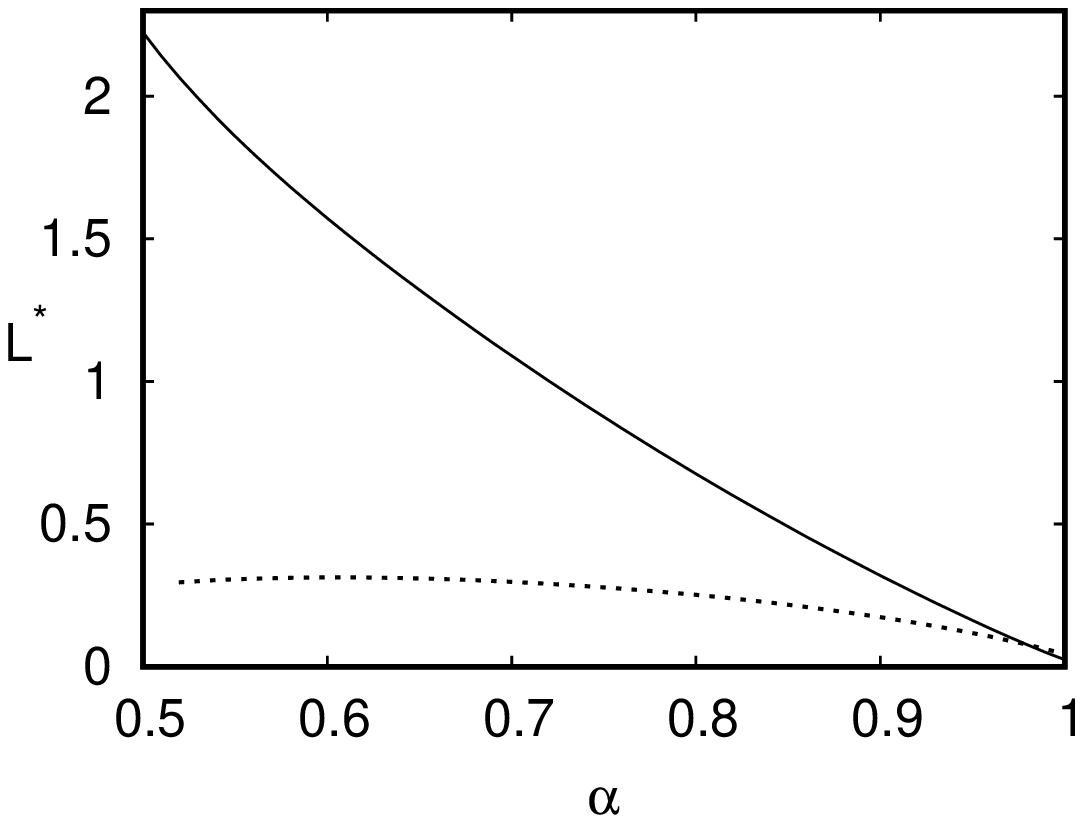}
\includegraphics[width=0.45\textwidth]{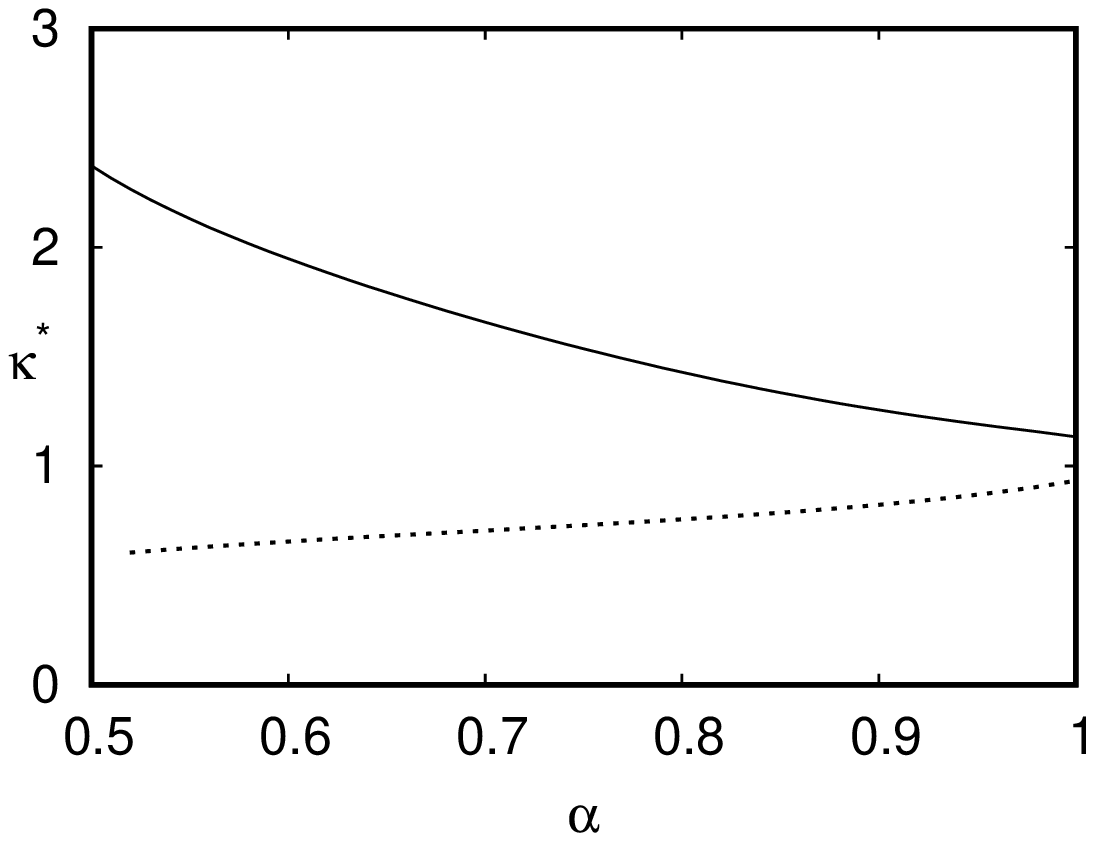}
\includegraphics[width=0.45\textwidth]{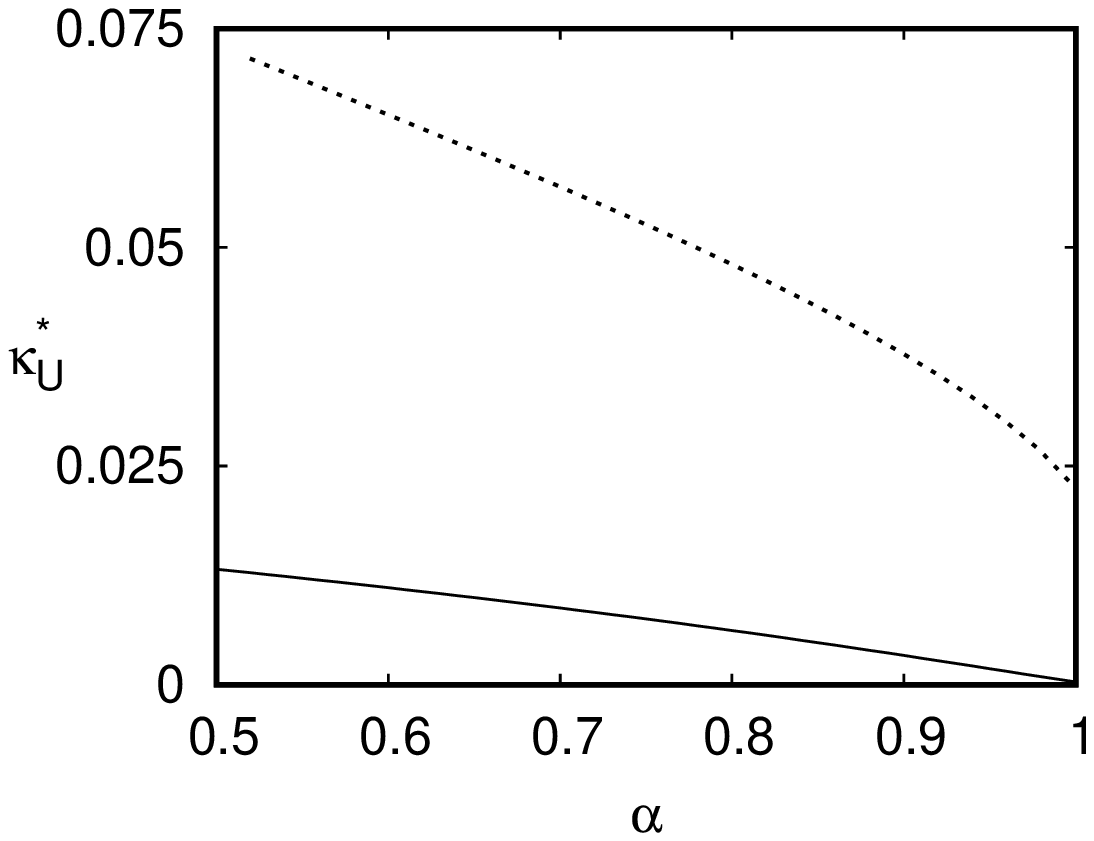}
\caption{Plot of the dimensionless transport coefficients associated with the heat flux as a function of the (common) coefficient of restitution $\alpha$ for driven granular mixtures under steady conditions ($\xi^*=\xi_{\text{st}}^*$). The parameters of the mixture are $d=2$, $m_1/m_2=2$, and $\sigma_1/\sigma_2=1$. The solid lines are for IMM while the dashed lines refer to IHS. The dimensionless coefficients $\vicentegp{D^{''*}}$ and $\kappa^*$ have been scaled with respect to their values for elastic collisions.}
\label{fig7}
\end{figure}
\begin{figure}[!h]
\centering
\includegraphics[width=0.45\textwidth]{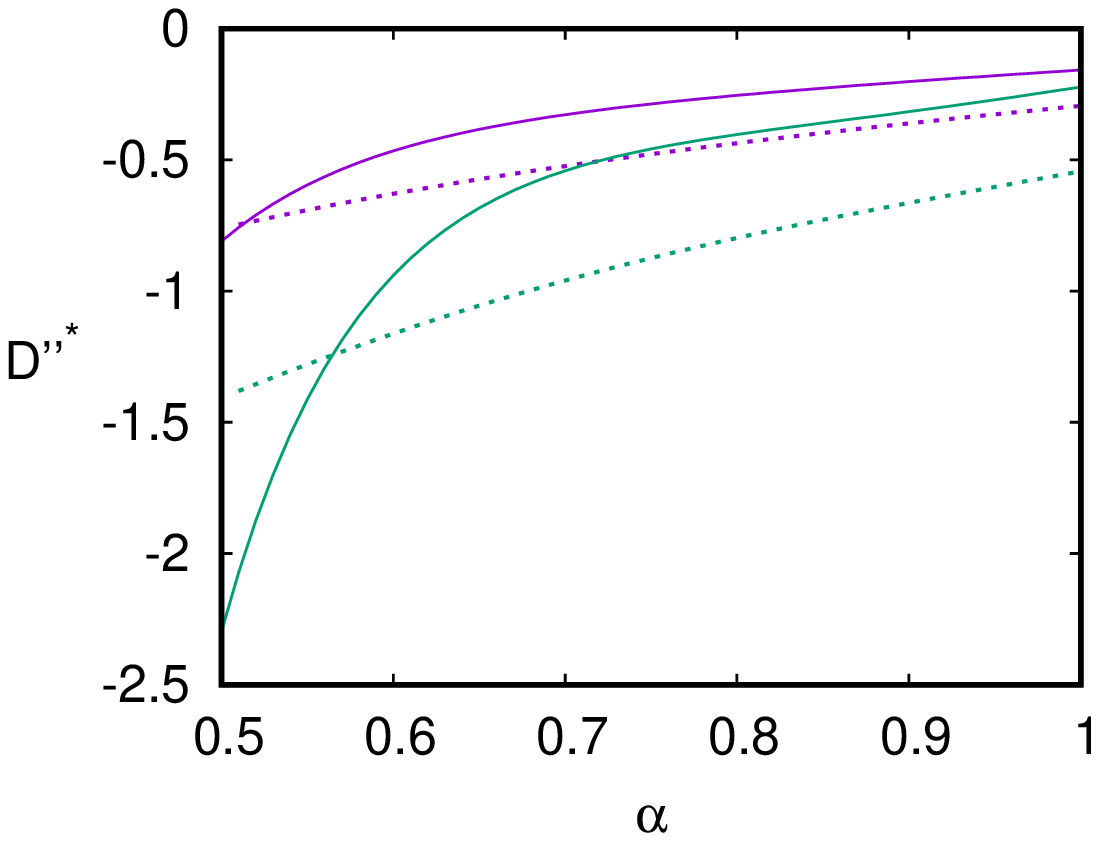}
\includegraphics[width=0.45\textwidth]{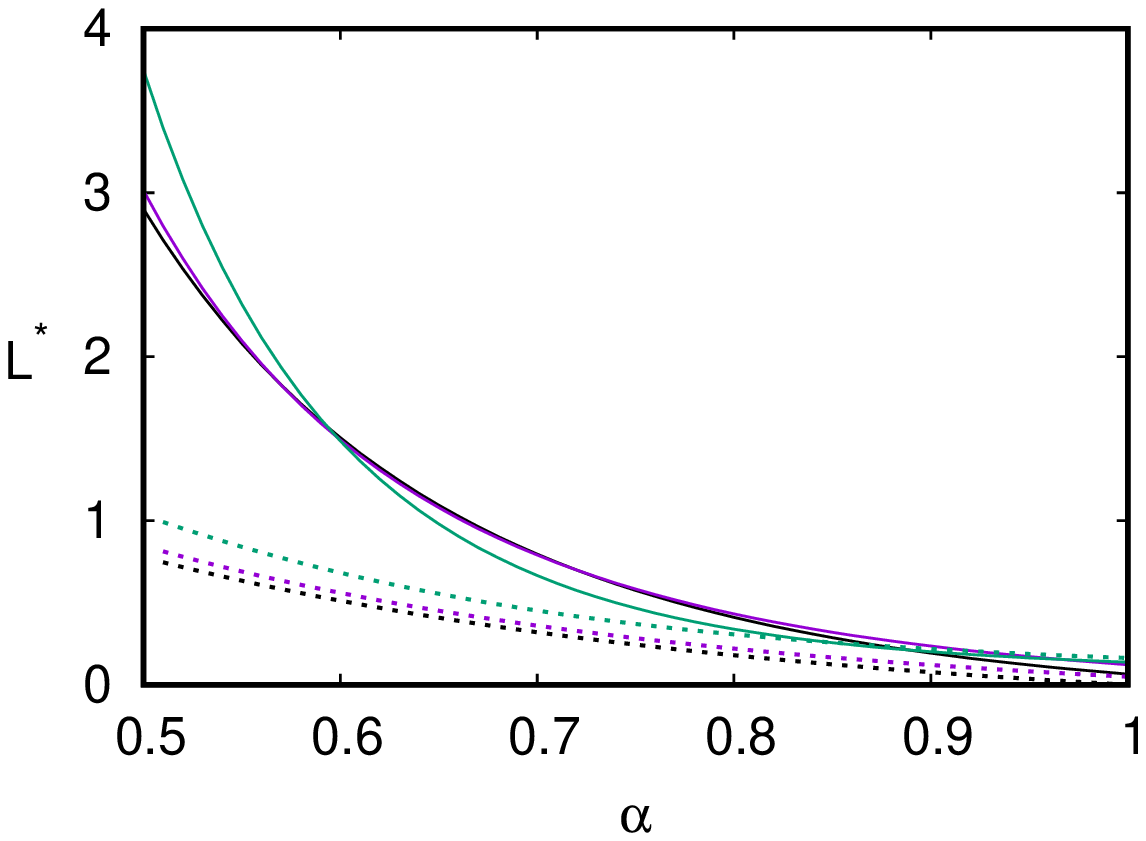}
\includegraphics[width=0.45\textwidth]{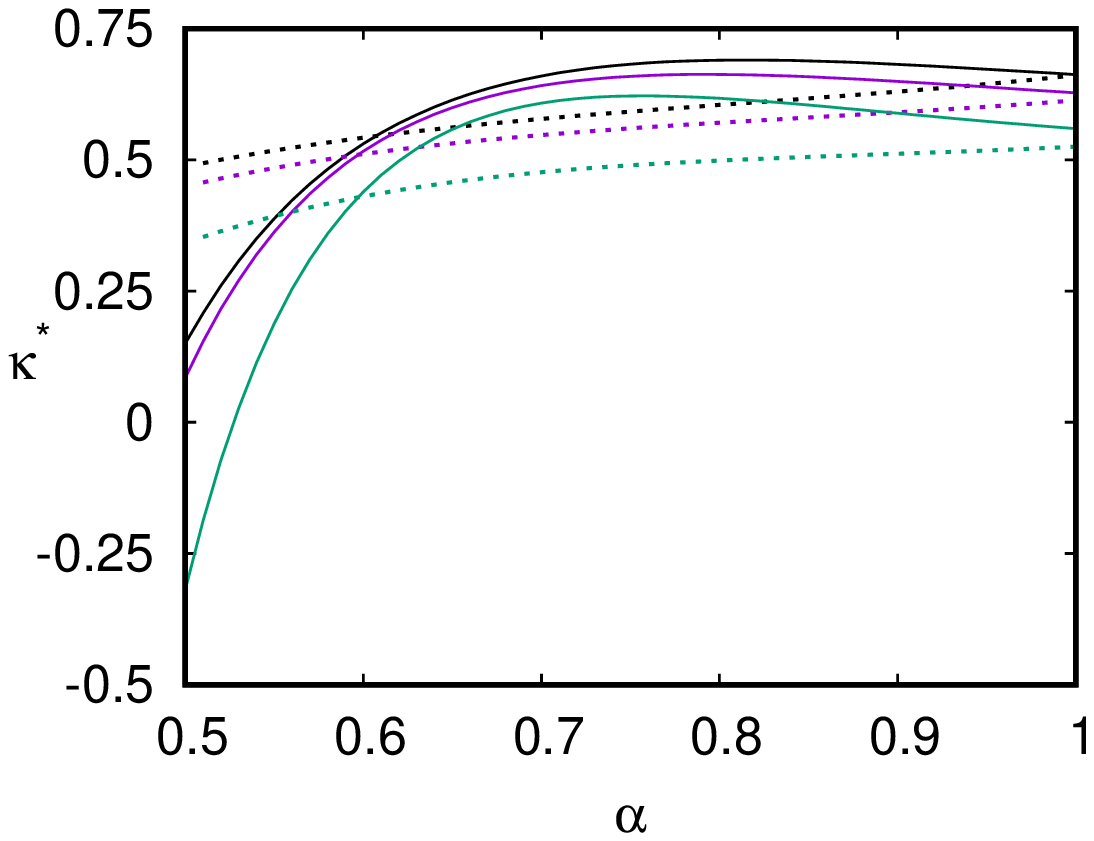}
\caption{Plot of the dimensionless transport coefficients associated with the heat flux as a function of the (common) coefficient of restitution $\alpha$ for undriven granular mixtures ($\xi^*=0$). The parameters of the mixture are $d=3$ and $\sigma_1/\sigma_2=1$; three different values of the mass ratio $m_1/m_2$ are considered: $m_1/m_2=1$ (black line), $m_1/m_2=2$ (violet line), and $m_1/m_2=4$ (green line). The solid lines refer to the results derived here for IMM while the dotted lines correspond to the results obtained for IHS in Ref.\ \cite{KG18}. The dimensionless coefficients $\vicentegp{D^{''*}}$ and $\kappa^*$ have been scaled with respect to their values for elastic collisions.}
\label{fig8}
\end{figure}

The $\al$--dependence of the coefficient $\chi_U$ of the first-order contribution to the partial temperatures is plotted in Fig.\ \ref{fig10} in the steady state ($\xi^*=\xi^*_{\text{st}}$) for a two-dimensional system with $x_1=\frac{1}{2}$, $\sigma_1=\sigma_2$, and two values of the mass ratio. We recall that this coefficient is zero for the undriven case ($\xi^*=0$). Given that the coefficient $\chi_U$ has not been determined so far for IHS, we cannot make any comparison between IMM and IHS for this transport coefficient. We observe that the magnitude of $\chi_U$ is not small, specially for high mass ratios and moderate inelasticity. This means that the first-order contribution to the partial temperatures cannot be neglected in the hydrodynamic description of the mixture (for instance, it should be taken into account in the stability analysis of the homogeneous state). Moreover, Fig.\ \ref{fig10} highlights that $\chi_U$ is positive and a decreasing function of $\alpha$. Regarding its dependence on the mass ratio, we observe that $\chi_U$ is an increasing function of $m_1/m_2$.

Figure \ref{fig6} plots the (scaled) shear viscosity coefficient $\eta^*(\al)/\eta^*(1)$ versus $\alpha$ for $d=2$, $x_1=\frac{1}{2}$, and $\sigma_1=\sigma_2$. Three different values of the mass ratio are considered.  As occurs in monocomponent granular gases \cite{S03,ChGV14}, we observe that in general the qualitative dependence of the Navier-Stokes shear viscosity of the mixture (for driven and undriven systems) on inelasticity of IHS is well reproduced by IMM: $\eta^*$ increases with decreasing $\al$. On the other hand, this increase is faster for IMM and so, the IMM predictions overestimate their IHS counterparts.

Finally, Figs.\ \ref{fig7} and \ref{fig8} show the dimensionless heat flux transport coefficients as a function of the coefficient of restitution for $\xi^*=\xi_{\text{st}}^*$ and $\xi^*=0$, respectively. Figure \ref{fig7} refers to hard disks ($d=2$) while Fig.\ \ref{fig8} corresponds to hard spheres ($d=3$). Although the theoretical results of IMM capture qualitatively the trends of IHS for some heat flux transport coefficients, significant quantitative discrepancies between both interaction models are found for strong inelasticity. These type of discrepancies were already reported for monocomponent granular gases \cite{S03,ChGV14}.

\section{Discussion}
\label{sec7}

This work has focused on the evaluation of the Navier--Stokes transport coefficients of a granular binary mixture driven by a stochastic bath with friction. The results have been obtained by solving the set of nonlinear (inelastic) Boltzmann equations by means of the Chapman--Enskog method \cite{CC70}. Since this method requires the choice of a reference base state (zeroth-order approximation $f_i^{(0)}$ in the perturbation expansion), as a first step we have characterized the time-dependent homogeneous state of the mixture. In particular, we have obtained the dependence of both the temperature ratio between the components of the mixture as well as the fourth cumulants (which measure the deviation of the distribution functions from their Maxwellian forms) on the (scaled) thermostat parameter $\xi^*(t)\propto T(t)^{-3/2}$ [$\xi^*$ being the reduced noise strength defined in Eq.\ \eqref{3.5})]. As a second step, we have derived the kinetic equation \eqref{4.14} verifying the first-order solution $f_i^{(1)} (\mathbf{v})$ to the Chapman--Enskog expansion. The knowledge of the distributions $f_i$ allowed us to determine the irreversible fluxes and identify the nine relevant Navier--Stokes transport coefficients of the mixture: four coefficients associated with the mass flux (the diffusion coefficient $D$, the pressure diffusion coefficient $D_p$, the thermal diffusion coefficient $D_T$, and the velocity diffusion coefficient $D_U$), the shear viscosity coefficient $\eta$ associated with the pressure tensor, and four coefficients associated with the heat flux (the Dufour coefficient $D''$, the pressure energy coefficient $L_p$, the thermal conductivity coefficient $D_T$, and the velocity conductivity coefficient $\kappa_U$).

On the other hand, it is important to remark that, unlike previous attempts for IHS \cite{KG13,KG18,KG19}, the present work considered a time-dependent reference state that can be far away from the homogeneous steady state. This means that the determination of the transport coefficients is not necessarily restricted to states near the above homogeneous states and so, the Navier--Stokes transport coefficients are in general given in terms of the (numerical) solution of a set of nonlinear differential equations. Analytical solutions to these equations can be obtained \emph{only} in two particular situations: (i) undriven granular mixtures ($\xi^*=0$) and (ii) driven mixtures in steady state conditions [$\xi^*=\xi_\text{st}^*$ where $\xi_\text{st}^*$ is obtained from the condition $x_1\Lambda_1^*+x_2\Lambda_2^*=0$, $\Lambda_i^*$ being defined by Eq.\ \eqref{3.9}]. Moreover, due to the technical difficulties involved in the time-dependent problem, we have considered here IMM instead of IHS to simplify the calculations and get the exact forms of the transport coefficients.
Regarding the homogeneous state, we have shown that the set of Boltzmann equations vicentegp{admits} the hydrodynamic scaling solutions \eqref{3.4} where the temperature dependence of the scaled distributions $\varphi_i$ occurs only through the (dimensionless) velocity $\mathbf{c}=\mathbf{v}/v_0(t)$ ($v_0(T)$ being the thermal speed) and the dimensionless noise strength $\xi^*(T)$. Although the exact form of the distributions $\varphi_i$ is not exactly known even for IMM, they can be characterized by their first velocity moments. In particular, we have studied the time evolution of the temperature ratio $\chi_1=T_1/T$ (which is formally equivalent to analyze the $\xi^*$--dependence of $\chi_1$) for different systems and different initial conditions. As figure \ref{fig1} clearly shows, after a short transient regime, all the curves collapse in an unsteady hydrodynamic solution $\chi_1(\xi^*)$ before reaching the asymptotic final steady state. The same behavior has been found for the fourth cumulants $K_i$ of the distributions $\varphi_i$ and similar time evolution is expected for higher cumulants.

Once the reference state is well characterized, the complete set of transport coefficients has been determined. As in the case of $\chi_1$ and $K_i$, we have seen that the (scaled) transport coefficients evolve in time towards the asymptotic steady state. Apart from the transport coefficients, we have also evaluated the first order contributions $T_i^{(1)}$ to the partial temperatures. These contributions are proportional to the divergence of the flow velocity (namely, $T_1^{(1)}=T \chi_U \nabla \cdot \mathbf{U}$ and $T_2^{(1)}=-(x_1/x_2)T \chi_U \nabla \cdot \mathbf{U}$). Although these coefficients are not hydrodynamic quantities, their calculation is interesting by itself and also because they are involved in the first order contribution $\zeta^{(1)}$ to the cooling rate. The existence of a nonzero first-order contribution $T_i^{(1)}$ induces a breakdown of the energy equipartition, additional to the one appearing in the homogeneous state (which is only due to the inelastic character of collisions). In fact, $T_i^{(1)}=0$ for undriven granular mixtures at low-density \cite{GD02} but $T_i^{(1)}\neq 0$ for moderately dense mixtures \cite{KS79b,GGG19b}. Although the existence of a non-vanishing contribution to the partial temperature $T_i^{(1)}$ for IHS has been recently recognized in an erratum \cite{KG19}, its expression for IHS has not been calculated so far. The results obtained in this paper for IMM show that the magnitude of the coefficient $\chi_U$ is in general not small and hence, the impact of $T_i^{(1)}$ on $\zeta^{(1)}$ cannot always be neglected.

Before considering the undriven and driven steady solutions, we have analyzed the time dependence of the (scaled) transport coefficients for given values of both the coefficients of restitution and the parameters of the mixture (masses, diameters, and concentration). This is in fact equivalent to studying the $\xi^*$--dependence of the (scaled) transport coefficients, which in turn allowed us to assess the influence of the thermostat on transport properties. As expected, for small inelasticity (say $\alpha \gtrsim 0.9$), the transport coefficients depend very weakly on $\xi^*$.  By contrast, the impact of $\xi^*$ on the (scaled) transport coefficients becomes in general more significant as the inelasticity increases. Thus, a very good approximation when describing driven IMM with small inelasticity is to use the expressions of the transport coefficients of the undriven case (keeping in mind that the constitutive equations have to include the terms of the thermostat). The previous conclusion is expected to be applicable to IHS as well.

As a complement of the previous results, we have also carried out an extensive comparison between the analytical expressions obtained here for IMM and those previously reported for undriven IHS mixtures \cite{GD02} and for IHS mixtures  driven by the same type of thermostat considered in this paper \cite{KG13,KG18}. To the best of our knowledge, this comparison between transport coefficients for granular mixtures of IMM and IHS had been only performed for the mass flux transport coefficients \cite{GA05} and for non-Newtonian transport in mixtures under uniform shear flow \cite{G03}. The comparison showed in general an excellent agreement between IMM and IHS for the transport coefficients associated with the mass flux (for both undriven and driven mixtures), a qualitative agreement for the shear viscosity coefficient, and significant quantitative discrepancies for the heat flux transport coefficients, specially at strong inelasticity.

\vicentegp{As a final comment, we want to emphasize that in this paper we have shown that a family of flow regimes which traditionally has been regarded as different when analysed through the Chapman--Enskog scheme can in fact be collected in a single group. This unification has been possible thanks to the use of a more general time-dependent reference state. This is, in our opinion, an important step towards having a unified hydrodynamic description of driven and undriven granular gases}.

\acknowledgments

The work of V.G. has been supported by the Spanish Government through Grant No. FIS2016-76359-P and by the Junta de Extremadura (Spain) Grant Nos. IB16013 and GR18079, partially financed by ``Fondo Europeo de Desarrollo Regional'' funds.

\appendix

\section{Some technical details on the evaluation of the transport coefficients}
\label{appA}

In this Appendix we provide some technical details on the calculation of the Navier--Stokes transport coefficients and the first-order contribution to the partial temperatures.

\subsection{Mass flux}

Let us start with the determination of the diffusion transport coefficients. The first-order contribution $\mathbf{j}_1^{(1)}$ to the mass flux is defined as
\beq
\label{5.2.1}
\mathbf{j}_1^{(1)}=\int\; \dd\mathbf{v}\; m_1 \mathbf{V} f_1^{(1)}(\mathbf{V}).
\eeq
To compute $\mathbf{j}_1^{(1)}$, we multiply both sides of Eq.\ \eqref{4.14} by $m_1 \mathbf{V}$ and integrate over velocity. After some algebra, we get
\beqa
\label{5.3}
\partial_t^{(0)}\mathbf{j}_1^{(1)}+\frac{\gamma_\text{b}}{m_1^\beta}\mathbf{j}_1^{(1)}+\nu_D \mathbf{j}_1^{(1)}&=&
-\left[p\frac{\partial}{\partial x_1}\left(x_1 \chi_1\right)+\frac{\gamma_\text{b} \rho_1 m_1 m_2 \delta m_\beta}{\rho^2 \overline m^\beta}\frac{p}{T} D \right]\nabla x_1 \nonumber\\
& & -\left[x_1 \left(\chi_1+p\frac{\partial \chi_1}{\partial p}\right)-\frac{\rho_1}{\rho}+\frac{\gamma_\text{b} \rho_1\delta m_\beta}{p \overline m^\beta}D_p \right]\nabla p \nonumber\\
& & -\left(px_1\frac{\partial \chi_1}{\partial T}+\frac{\gamma_\text{b} \rho_1\delta m_\beta}{\overline m^{\beta}}D_T\right) \nabla T
-\frac{\gamma_\text{b}\rho_1}{\rho} \frac{\delta m_\beta}{\overline m^{\beta}}\left(\rho_2+ D_U\right)\Delta \mathbf{U}.
\eeqa
Upon obtaining Eq.\ \eqref{5.3}, use has been made of the result \cite{GA05}
\beq
\label{5.4}
\int\; \dd\mathbf{v}\; m_1 \mathbf{V}\left(\mathcal{L}_1f_1^{(1)}+\mathcal{M}_1f_2^{(1)}\right)=\nu_D \mathbf{j}_1^{(1)},
\eeq
where
\beq
\label{5.5}
\nu_D=\rho \frac{\nu_{12}}{dn_2}\frac{1+\al_{12}}{m_1+m_2}.
\eeq
The solution to Eq.\ \eqref{5.3} is of the form \eqref{4.13}, as expected. Dimensional analysis shows that $D\propto T^{1/2}$, $D_p\propto D_T\propto T^{3/2}/p$, and $D_U \propto p/T$ and hence, $\partial_t^{(0)}\left\{D, D_p, D_T, D_U\right\}=-\frac{1}{2}\Lambda^{(0)}\left\{D, D_p, D_T, 0\right\}$. Thus, the time derivative $\partial_t^{(0)} \mathbf{j}_1^{(1)}$ can be computed as
\beqa
\label{5.6}
\partial_t^{(0)}\mathbf{j}_1^{(1)}&=& \frac{p}{\nu_0}\left[\Lambda^{(0)}\left(\frac{1}{2}D^*-\frac{3}{2}\xi^*\partial_{\xi^*} D^*\right)+(D_p^*+D_T^*)\partial_{x_1} \Lambda^{(0)}\right]\nabla x_1+\frac{1}{\nu_0}\Bigg[\Lambda^{(0)}\left(\frac{1}{2}D_p^*-\frac{3}{2}\xi^*\partial_{\xi^*} D_p^*\right)\nonumber\\
& &+(D_p^*+D_T^*)p\partial_p \Lambda^{(0)}\Bigg]\nabla p+\frac{p}{T\nu_0}\left[\Lambda^{(0)}\left(\frac{1}{2}D_T^*-\frac{3}{2}\xi^*\partial_{\xi^*} D_T^*\right)+(D_p^*+D_T^*)T\partial_T \Lambda^{(0)}\right]\nabla T\nonumber\\
& &-\frac{3p\overline m\Lambda^{(0)}}{2T}\left(\xi^*\partial_{\xi^*} D_U^*\right)\Delta \mathbf U,
\eeqa
where we have introduced the dimensionless coefficients
\beq
\label{5.7}
D=\frac{\rho T}{m_{1}m_{2}\nu_{0}}D^*,\quad
D_{p}=\frac{p}{\rho\nu_{0}}D_{p}^*,\quad
D_T=\frac{p}{\rho \nu_{0}}D_T^*, \quad D_U=\frac{p\overline m}{T}D_U^*.
\eeq
The diffusion coefficients $D$, $D_p$, $D_T$, and $D_U$ can be easily identified after inserting Eq.\ \eqref{5.7} into Eq.\ \eqref{5.3}. While the (reduced) coefficients $D^*$, $D_p^*$, and $D_T^*$ obey a set of coupled differential equations, the (reduced) coefficient $D_U^*$ obeys an autonomous equation,
\beq
\label{5.8}
\frac{3\Lambda^{(0)}}{2\nu_0}\xi^*\partial_{\xi^*}D_U^*+a_{44}D_U^*=a_{40},
\eeq
where the coefficients $a_{ij}$ are defined in Appendix \ref{appB}. In matrix form, the remaining coefficients verify the following set of differential equations:
\beq
\label{5.9}
\left(
\begin{array}{ccc}
a_{11}+\frac{3\Lambda^{(0)}}{2\nu_0}\xi^*\partial_\xi^*&a_{12}&a_{12}\\
0&a_{22}+\frac{3\Lambda^{(0)}}{2\nu_0}\xi^*\partial_\xi^*&a_{23}\\
0&a_{32}&a_{33}+\frac{3\Lambda^{(0)}}{2\nu_0}\xi^*\partial_\xi^*
\end{array}
\right)
\left(
\begin{array}{c}
D^*\\
D_p^*\\
D_T^*
\end{array}
\right)=
\left(
\begin{array}{c}
a_{10}\\
a_{20}\\
a_{30}
\end{array}
\right).
\eeq

Note that there are two ways of ``removing'' the presence of the derivatives $\partial_\xi^*$ in Eqs. \eqref{5.8} and \eqref{5.9}: (i) either by taking the limit $\Lambda^{(0)}\to 0$ (the system and the thermostat locally thermalize and a steady state is achieved) or (ii) by taking the limit $\xi^*\to 0$ (undriven granular mixtures). The former limit was analyzed in Refs. \cite{KG13,KG18} for IHS, while the latter was studied in Ref.\ \cite{GD02} for IHS and in Ref.\ \cite{GA05} for IMM. In both limit situations ($\Lambda^{(0)}=0$ or $\xi^*\to 0$), we can obtain analytical expressions for the diffusion transport coefficients. However, beyond both special situations, as expected we have to get the above coefficients by numerically solving Eqs.\ \eqref{5.8} and \eqref{5.9}.

\subsection{Pressure tensor}

The first-order contribution $\mathsf{P}^{(1)}$ to the pressure tensor can be written as $\mathsf{P}^{(1)}=\mathsf{P}_1^{(1)}+\mathsf{P}_2^{(1)}$, where
\begin{equation}
\label{5.10}
\mathsf P^{(1)}_i=m_i\int \dd\mathbf v\ \mathbf V\mathbf V f_i^{(1)}(\mathbf v).
\end{equation}
The partial contributions $\mathsf P^{(1)}_1$ can be obtained by multiplying both sides of Eq.\ \eqref{4.14} by $m_1 \mathbf{V}\mathbf{V}$ and integrating over $\mathbf{V}$. After some algebra, we have
\begin{eqnarray}
\label{5.11}
\partial_t^{(0)}P_{1,k\ell}^{(1)}&& +\left(\frac{2\gamma_b}{m_1^\beta}+\tau_{11}\right)\mathsf P_{1,k\ell}^{(1)}+\tau_{12}P_{2,k\ell}^{(1)}=-p_1^{(0)}\left(\partial_k U_\ell+\partial_\ell U_k-\frac{2}{d}\delta_{k\ell}\nabla\cdot \mathbf U\right)\\ \nonumber && +\Bigg[p \partial_p p_1^{(0)}-\frac{d+2}{d}p_1^{(0)}+\left(\frac{2}{d}+\zeta_U+2\gamma_b x_1\frac{\delta m_\beta}{\overline m^\beta}\chi_U\right)\left(p\partial_p+T\partial_T\right)p_1^{(0)}\Bigg]\delta_{k\ell}\nabla\cdot \mathbf U,
\end{eqnarray}
where $p_1^{(0)}=n_1 T_1^{(0)}$ and use has been made of the result \cite{GA05}
\begin{equation}
\label{5.12}
\int \dd\mathbf v \ m_1 \mathbf V\mathbf V \left({\mathcal L}_{1} f_{1}^{(1)} +{\mathcal M}_{1}f_{2}^{(1)}\right)=\tau_{11}\mathsf P_1^{(1)}+\tau_{12}\mathsf P_2^{(1)},
\end{equation}
where
\begin{equation}
\label{5.13}
\tau_{11}=\frac{\nu_{11}}{d(d+2)}(1+\alpha_{11})(d+1-\alpha_{11})+2\frac{\nu_{12}}{d}\mu_{21}
(1+\alpha_{12})\left[1-\frac{\mu_{21}(1+\alpha_{12})}{d+2}\right], \quad \tau_{12}=-2\frac{\nu_{12}}{d(d+2)}\frac{\rho_1}{\rho_2}\mu_{21}^2
(1+\alpha_{12})^2.
\end{equation}
The corresponding equation for $\mathsf P_2^{(1)}$ is
\begin{eqnarray}
\label{5.13.0}
\partial_t^{(0)}P_{2,k\ell}^{(1)}&& +\left(\frac{2\gamma_b}{m_2^\beta}+\tau_{22}\right)\mathsf P_{2,k\ell}^{(1)}+\tau_{21}P_{1,k\ell}^{(1)}=-p_2^{(0)}\left(\partial_k U_\ell+\partial_\ell U_k-\frac{2}{d}\delta_{k\ell}\nabla\cdot \mathbf U\right)\\ \nonumber && +\Bigg[p \partial_p p_2^{(0)}-\frac{d+2}{d}p_2^{(0)}+\left(\frac{2}{d}+\zeta_U+2\gamma_b x_1\frac{\delta m_\beta}{\overline m^\beta}\chi_U\right)\left(p\partial_p+T\partial_T\right)p_2^{(0)}\Bigg]\delta_{k\ell}\nabla\cdot \mathbf U.
\end{eqnarray}
The expressions of the collision frequencies $\tau_{22}$ and $\tau_{21}$ can be taken from Eq.\ \eqref{5.13} after interchanging $1\leftrightarrow 2$.

Contrary to what happens in the undriven case \cite{GD02,GA05}, Eqs.\ \eqref{5.11} and \eqref{5.13.0} show clearly that $\text{Tr} \mathsf P_i^{(1)}=d p_i^{(1)}=d n_i T_i^{(1)}\neq 0$. The equation defining the first-order contribution $p_1^{(1)}$ to the partial pressure of component $1$ can be easily derived by taking the trace in Eq.\ \eqref{5.11} or, alternatively, by multiplying Eq. \eqref{4.14} by $m_1V^2$ and integrating over $\mathbf{v}$. The result is
\beq
\label{5.13.1}
\partial_t^{(0)}p_1^{(1)} + \left(\frac{2\gamma_b}{m_1^\beta}+\tau_{11}\right)p_1^{(1)}+\tau_{12}p_2^{(1)}=-D_t^{(1)}p_1^{(0)}-\frac{d+2}{2} p_1^{(0)}\nabla \cdot \mathbf{U}.
\eeq
The corresponding equation for $p_2^{(1)}$ can be easily obtained from Eq.\ \eqref{5.13.1} by the change $1\leftrightarrow 2$. Summing the equations for $p_1^{(1)}$ and $p_2^{(2)}$, we find that $p_1^{(1)}=-p_2^{(1)}$, in accordance with the consistency condition defined in the second relation of Eq.\ \eqref{eq:pr1}. This means that the granular temperature $T$ is not affected by the spatial gradients.

%To prove the constraint $p_1^{(1)}=-p_2^{(1)}$, we need to take into account the relations $x_1\chi_1+x_2\chi_2=1$, $p_1^{(0)}+p_2^{(0)}=p$, and $\zeta_U=x_1 (\tau_{11}-\tau_{12}-\tau_{22}+\tau_{21})$.

Equation \eqref{5.13.1} has the solution $p_1^{(1)}=x_1\frac{p}{\nu_0}\chi_U\nabla \cdot \mathbf{U}$, where $\chi_U$ verifies
\begin{eqnarray}
\label{5.13.2}
\nonumber \frac{3\Lambda^*}{2} \xi^*\partial_{\xi^*}\chi_U&+&\left[-\frac{\Lambda^*}{2}+\frac{2\omega^*\xi^{*1/3}}{M_1^\beta}+\tau_{11}^*-\tau_{12}^*+ \left(\tau_{11}^*+\tau_{21}^*-\tau_{22}^*-\tau_{12}^*+2\omega^*\xi^{*1/3}\delta m_{\beta}\right)\frac{3x_1}{2} \xi^*\partial_{\xi^*}\chi_1\right]\chi_U \\ && =-\frac{d+3}{d}\xi^*\partial_{\xi^*}\chi_1-\frac{2}{3}\omega^*\partial_{\omega^*}\chi_1,
\end{eqnarray}
where $\tau_{ij}^*=\tau_{ij}/\nu_0$, and use has been made of the relations $p \partial_p \chi_1= -\xi^*\partial_\xi^* \chi_1-(2/3)\omega^*\partial_\omega^* \chi_1$ and $T\partial_T \chi_1= -(1/2)\xi^*\partial_\xi^* \chi_1+(2/3)\omega^*\partial_\omega^* \chi_1$.
Note that for $\xi^*\to 0$, Eq.\ \eqref{3.12} yields $\omega^*\partial_{\omega^*}\chi_1\to 0$ and so, Eq.\ \eqref{5.13.2} leads to
$\chi_U=0$ as expected \cite{GD02}. However, when $\xi^*\ne 0$, the right hand side of Eq.\ \eqref{5.13.2} is in general different from zero and hence $\chi_U\ne 0$ for driven granular mixtures at low density.

To identify the shear viscosity coefficient $\eta$, it is convenient to rewrite $P_{1,k\ell}^{(1)}$ as
\beq
\label{5.14}
P_{1,k\ell}^{(1)}=p_1^{(1)}\delta_{k\ell}+\Pi_{1,k\ell}^{(1)},
\eeq
where $\Pi_{1,k\ell}^{(1)}$ is the traceless part of the partial pressure tensor $P_{1,k\ell}^{(1)}$. From Eq.\ \eqref{5.11}, we get the differential equation obeying $\Pi_{1,k\ell}^{(1)}$:
\beq
\label{5.15}
\partial_t^{(0)}\Pi_{1,k\ell}^{(1)}+\left(\frac{2\gamma_b}{m_1^\beta}+\tau_{11}\right)\Pi_{1,k\ell}^{(1)}+\tau_{12}\Pi_{1,k\ell}^{(1)}
=-px_1\chi_1\left(\partial_k U_\ell+\partial_\ell U_k-\frac{2}{d}\delta_{k\ell}\nabla\cdot \mathbf U\right).
\eeq
The differential equation of $\Pi_{2,k\ell}^{(1)}$ can be easily inferred from Eq.\ \eqref{5.15} by interchanging $1\leftrightarrow 2$. The solution to Eq.\ \eqref{5.15} (and its counterpart for $\Pi_{2,k\ell}^{(1)}$) can be written as
\begin{equation}
\label{5.16}
\Pi_{i,k\ell}^{(1)}=-\eta_i\left(\frac{\partial U_k}{\partial r_\ell}+\frac{\partial U_\ell}{\partial r_k}-\frac{2}{d}\delta_{k\ell}\nabla \cdot \mathbf{U}\right), \quad i=1,2.
\end{equation}
According to Eq.\ \eqref{5.1}, the shear viscosity of the mixture is $\eta=\eta_1+\eta_2$. Dimensional analysis requires that $\eta_i\propto T^{1/2}$ and so,
\beq
\label{5.17}
\partial_t^{(0)}\eta_i=-\frac{p}{\nu_0}\Lambda^{(0)}\left(\frac{1}{2}\eta_i^*-\frac{3}{2}\xi^* \frac{\partial \eta_i^*}{\partial  \xi^*}\right),
\eeq
where $\eta_i^*=(\nu_0/p)\eta_i$. Thus, in matrix form, the set of equations for $\eta^*_i$ is given by
\beq
\label{5.18}
\left(
\begin{array}{cc}
b_{11}+\frac{3\Lambda^{(0)}}{2\nu_0}\xi^*\partial_\xi^*&b_{12}\\
b_{21}&b_{22}+\frac{3\Lambda^{(0)}}{2\nu_0}\xi^*\partial_\xi^*
\end{array}
\right)
\left(
\begin{array}{c}
\eta_1^*\\
\eta_2^*
\end{array}
\right)=
\left(
\begin{array}{c}
b_{10}\\
b_{20}
\end{array}
\right),
\eeq
where the coefficients $b_{ij}$ are defined in Appendix \ref{appB}. The solution to Eq.\ \eqref{5.18} gives the shear viscosity coefficient $\eta$. In the case of undriven granular gases ($\xi^*\to 0$), Eq.\ \eqref{5.18} agrees with the one derived before for IMM \cite{GA05}. Moreover, for steady state conditions ($\Lambda^{(0)}=0$), we also obtain a simple analytical solution. Beyond both limit cases, the numerical solution to the set of equations \eqref{5.18} provides the shear viscosity coefficient in the time-dependent driven state.

\subsection{Heat flux}

To first order, the heat flux is given by
\begin{equation}
\label{b1}
\mathbf q^{(1)}=-T^2D^{\prime \prime}\nabla x_1-L\nabla p-\kappa\nabla T-\kappa_{U}\Delta\mathbf U,
\end{equation}
where, in dimensionless forms, the Dufour coefficient $D^{\prime \prime}$, the pressure energy coefficient $L$, the thermal conductivity $\kappa$, and the velocity conductivity $\kappa_U$ are defined as
\beq
\label{b2}
D^{\prime \prime}=\frac{p}{T\overline m\nu_0}\left(D_1^{\prime \prime *}+D_2^{\prime \prime *}\right), \quad L=\frac{T}{\overline m\nu_0}\left(L_1^*+L_2^*\right), \quad
\kappa=\frac{p}{\overline m\nu_0}\left(\kappa_1^*+\kappa_2^*\right), \quad \kappa_U=p\left(\kappa_{U1}^*+\kappa_{U2}^*\right).
\eeq
The differential equations verifying the (scaled) coefficients $D_i^{\prime*}$, $L_i^*$, $\kappa_i^*$, and $\kappa_{Ui}^*$ can be obtained by following similar mathematical steps as those made for the other transport coefficients. As in the case of the diffusion coefficients, the (reduced) coefficients $\kappa_{Ui}^*$ verify an autonomous set of equations given by
\beq
\label{b3}
\left(
\begin{array}{cc}
c_{77}+\frac{3\Lambda^{(0)}}{2\nu_0}\xi^*\partial_\xi^*&c_{78}\\
c_{87}&c_{88}+\frac{3\Lambda^{(0)}}{2\nu_0}\xi^*\partial_\xi^*
\end{array}
\right)
\left(
\begin{array}{c}
\kappa_{U1}^*\\
\kappa_{U2}^*
\end{array}
\right)=
\left(
\begin{array}{c}
c_{70}\\
b_{80}
\end{array}
\right),
\eeq
where the expressions of the coefficients $c_{ij}$ are displayed in Appendix \ref{appB}. The remaining coefficients are coupled. By using matrix notation, the coupled set of six differential equations for the unknowns
\beq
\label{b4}
\left\{D_1^{''*}, D_2^{''*}, L_1^*, L_2^*, \kappa_1^*, \kappa_2^*\right\}
\eeq
can be written as
\beq
\label{b5}
\Sigma_{\mu\nu}X_\nu=Y_\mu.
\eeq
Here, $X_\nu$ is the column matrix defined by the set \eqref{b4}, $\Sigma_{\mu\nu}$ is the square matrix
\beq
\label{b6}
\left(
\begin{array}{cccccc}
c_{11}+\frac{3\Lambda^{(0)}}{2\nu_0}\xi^*\partial_\xi^*&c_{12}&c_{13}&0&c_{13}&0\\
c_{21}&c_{22}+\frac{3\Lambda^{(0)}}{2\nu_0}\xi^*\partial_\xi^*&0&c_{24}&0&c_{24}\\\
0&0&c_{33}+\frac{3\Lambda^{(0)}}{2\nu_0}\xi^*\partial_\xi^*&c_{34}&c_{35}&0\\
0&0&c_{43}&c_{44}+\frac{3\Lambda^{(0)}}{2\nu_0}\xi^*\partial_\xi^*&0&c_{46}\\
0&0&c_{53}&0&c_{55}+\frac{3\Lambda^{(0)}}{2\nu_0}\xi^*\partial_\xi^*&c_{56}\\
0&0&0&c_{64}&c_{65}&c_{66}+\frac{3\Lambda^{(0)}}{2\nu_0}\xi^*\partial_\xi^*
\end{array}
\right),
\eeq
and the column matrix $\mathsf{Y}$ is
\beq
\label{b7}
\mathsf{Y}=\left(
\begin{array}{c}
c_{10}\\
c_{20}\\
c_{30}\\
c_{40}\\
c_{50}\\
c_{60}
\end{array}
\right).
\eeq
In the undriven ($\xi^*=0$) and driven steady states ($\xi^*=\xi_\text{st}^*$) the solution to Eq.\ \eqref{b5} can be written as
\beq
\label{b8}
X_\mu=(\Sigma)^{-1}_{\mu\nu}Y_\nu.
\eeq

\section{Expressions of the coefficients $a_{ij}$, $b_{ij}$,  and $c_{ij}$ \label{appB}}

In this Appendix we display the explicit expressions of the coefficients $a_{ij}$, $b_{ij}$, and $c_{ij}$ defining the diffusion coefficients, the shear viscosity coefficient, and the heat flux coefficients, respectively.

The coefficients $a_{ij}$ are introduced in Eqs.\ \eqref{5.8} and \eqref{5.9} for the evaluation of the (reduced) diffusion transport coefficients $D_U^*$, $D^*$, $D_p^*$, and $D_T^*$. They are given by
\beq
\label{a1}
% a_{11} esta mal en khga13 porque \nu_D debe estar dividido por \nu_0
a_{10}=\partial_{x_{1}}(x_{1}\chi_{1}), \quad a_{11}=-\frac{\Lambda^{(0)}}{2\nu_0}+\frac{\nu_D}{\nu_0}+\omega^* \xi^{*1/3}\frac{\rho_1 m_1^\beta+\rho_2 m_2^\beta}{\rho (m_1+m_2)^\beta}, \quad a_{12}=-\frac{1}{\nu_0}\partial_{x_1} \Lambda^{(0)},
\eeq
\beq
\label{a2}
a_{20}=x_1\chi_1-\frac{\rho_1}{\rho}+x_1 p \partial_p \chi_1, \quad  a_{22}=a_{11}+a_{23}, \quad a_{23}=-\frac{p}{\nu_0}\partial_p \Lambda^{(0)},
\eeq
\beq
\label{a3}
a_{30}= x_1 T \partial_T \chi_1, \quad a_{32}=-\frac{T}{\nu_0}\partial_T \Lambda^{(0)},\quad   a_{33}=a_{11}+a_{32},
\eeq
\beq
\label{a4}
a_{40}=\frac{T\rho_1\rho_2}{p\overline m\rho }\omega^* \xi^{*1/3}\delta m_\beta, \quad a_{44}=\frac{\nu_D}{\nu_0}+\omega^* \xi^{*1/3}\frac{\rho_1 m_1^\beta+\rho_2 m_2^\beta}{\rho (m_1+m_2)^\beta},
\eeq
where
\beq
\label{a5}
\frac{1}{\nu_0}\partial_{x_1} \Lambda^{(0)}=2\omega^*\xi^{*1/3} \delta m_\beta\partial_{x_1}(x_1\chi_1)-\xi^{*} \delta m_{\lambda-1}+\partial_{x_1} \zeta^*, \quad
\frac{p}{\nu_0}\partial_p \Lambda^{(0)}=2\omega^*\xi^{*1/3} \delta m_\beta x_1p\partial_p \chi_1+\frac{p}{\nu_0}\partial_p \zeta^*,
\eeq
\beq
\label{a6}
\frac{T}{\nu_0}\partial_T \Lambda^{(0)}=2\omega^*\xi^{*1/3} \delta m_\beta x_1T\partial_T \chi_1+\xi^*\sum_{i=1}^2\frac{x_i}{M_i^{\lambda-1}}+\frac{T}{\nu_0}\partial_T \zeta^*.
\eeq

The coefficients $b_{ij}$ defining the shear viscosity in Eq. \eqref{5.18} are
\begin{eqnarray}
  &&b_{10}=x_1\chi_1, \qquad b_{11}=\frac{\tau_{11}}{\nu_0}+\frac{2\omega^*\xi^*}{M_1^\beta}-\frac{\Lambda^{(0)}}{2\nu_0}, \qquad b_{12}=\frac{\tau_{12}}{\nu_0},\\
  &&b_{20}=x_2\chi_2, \qquad b_{21}=\frac{\tau_{21}}{\nu_0}, \qquad b_{22}=\frac{\tau_{22}}{\nu_0}+\frac{2\omega^*\xi^*}{M_2^\beta}-\frac{\Lambda^{(0)}}{2\nu_0},
\end{eqnarray}

The coefficients $c_{ij}$ of the heat flux are
\beq
\label{b8}
c_{10}=\left[-\frac{\overline m\epsilon_{12}}{T\nu_0}+(d+2)\frac{\xi^*}{2M_1^\lambda}+(d+2)\frac{\omega^*\xi^{*1/3}}{2M_1}\frac{\delta m_\beta m_1x_1\chi_1}{x_1m_1+x_2m_2}\right]D^*+\frac{d+2}{2M_1}\partial_{x_1}\left[\left(1+\frac{K_1}{2}\right)
x_1\chi_1^{2}\right],
\nonumber
\eeq
\beq
\label{b9}
c_{11}=-\frac{3\Lambda^{(0)}}{2\nu_0}+\frac{\beta_{11}}{\nu_0}+\frac{3\omega^*\xi^{*1/3}}{M_1^\beta}, \quad c_{12}=\frac{\beta_{12}}{\nu_0}, \quad c_{13}=-\frac{1}{\nu_0}\partial_{x_1}\Lambda^{(0)},
\eeq
\beq
\label{b10}
c_{20}=\left[\frac{\overline m\epsilon_{21}}{T\nu_0}-(d+2)\frac{\xi^*}{2M_2^\lambda}-(d+2)\frac{\omega^*\xi^{*1/3}}{2M_2}
\frac{\delta m_\beta m_2x_2\chi_2^{(0)}}{x_1m_1+x_2m_2}\right]D^*+\frac{d+2}{2M_2}\partial_{x_1}\left[\left(1+\frac{K_2}{2}\right)
x_2\chi_2^{2}\right],
\eeq
\beq
\label{b11}
c_{21}=\frac{\beta_{21}}{\nu_0}, \quad c_{22}=-\frac{3\Lambda^{(0)}}{2\nu_0}+\frac{\beta_{22}}{\nu_0}+\frac{3\omega^*\xi^{*1/3}}{M_2^\beta}, \quad c_{24}=-\frac{1}{\nu_0}\partial_{x_1}\Lambda^{(0)},
\eeq
\beqa
\label{b12}
c_{30}&=&\left[-\frac{\overline m \epsilon_{12}}{T\nu_0}+(d+2)\frac{\xi^*}{2M_1^\lambda}+(d+2)\frac{\omega^*\xi^{*1/3}}{2M_1}\frac{\delta m_\beta m_1x_1\chi_1}{x_1m_1+x_2m_2}\right]D_p^* +\frac{d+2}{2M_1}\Bigg\{\partial_{p}\left[p\left(1+\frac{K_1}{2}\right)x_1\chi_1^2\right]
\nonumber\\
& &
-\frac{m_1x_1\chi_1}{x_1m_1+x_2m_2}\Bigg\},
\eeqa
\beq
\label{b13}
c_{33}=-\frac{3\Lambda^{(0)}}{2\nu_0}-\frac{1}{\nu_0}p\partial_p\Lambda^{(0)}+\frac{\beta_{11}}{\nu_0}
+\frac{3\omega^*\xi^{*1/3}}{M_1^\beta}, \quad c_{34}=\frac{\beta_{12}}{\nu_0}, \quad c_{35}=-\frac{1}{\nu_0}p\partial_{p}\Lambda^{(0)},
\eeq
\beqa
\label{b14}
c_{40}&=&\left[\frac{\overline m \epsilon_{21}}{T\nu_0}-(d+2)\frac{\xi^*}{2M_2^\lambda}-(d+2)\frac{\omega^*\xi^{*1/3}}{2M_2}\frac{\delta m_\beta m_2x_2\chi_2}{x_1m_1+x_2m_2}\right]D_p^* +\frac{d+2}{2M_2}\Bigg\{\partial_{p}\left[p\left(1+\frac{K_2}{2}\right)x_2\chi_2^2\right]
\nonumber\\
& &
-\frac{m_2x_2\chi_2}{x_1m_1+x_2m_2}\Bigg\},
\eeqa
\beq
\label{b15}
c_{43}=\frac{\beta_{21}}{\nu_0}, \quad c_{44}=-\frac{3\Lambda^{(0)}}{2\nu_0}-\frac{1}{\nu_0}p\partial_p\Lambda^{(0)}+\frac{\beta_{22}}{\nu_0}
+\frac{3\omega^*\xi^{*1/3}}{M_2^\beta}, \quad c_{46}=-\frac{1}{\nu_0}p\partial_{p}\Lambda^{(0)},
\eeq
\beq
\label{b16}
c_{50}=\left[-\frac{\overline m\epsilon_{12}}{T\nu_0}+(d+2)\frac{\xi^*}{2M_1^\lambda}+(d+2)\frac{\omega^*\xi^{*1/3}}{2M_1}\frac{\delta m_\beta m_1x_1\chi_1}{x_1m_1+x_2m_2}\right]D_T^*+\frac{d+2}{2M_1}\partial_{T}\left[T\left(1+\frac{K_1}{2}\right)
x_1\chi_1^2\right], \nonumber\\
\eeq
\beq
\label{b17}
c_{53}=-\frac{1}{\nu_0}T\partial_{T}\Lambda^{(0)}, \quad c_{55}=-\frac{3\Lambda^{(0)}}{2\nu_0}-\frac{1}{\nu_0}T\partial_T\Lambda^{(0)}+\frac{\beta_{11}}{\nu_0}
+\frac{3\omega^*\xi^{*1/3}}{M_1^\beta}, \quad c_{56}=\frac{\beta_{12}}{\nu_0},
\eeq
\beq
\label{b18}
c_{60}=\left[\frac{\overline m\epsilon_{21}}{T\nu_0}-(d+2)\frac{\xi^*}{2M_2^\lambda}-(d+2)\frac{\omega^*\xi^{*1/3}}
{2M_2}\frac{\delta m_\beta m_2x_2\chi_2}{x_1m_1+x_2m_2}\right]D_T^*+\frac{d+2}{2M_2}\partial_{T}
\left[T\left(1+\frac{K_2}{2}\right)x_2\chi_2^2\right],
\eeq
\beq
\label{b19}
c_{64}=-\frac{1}{\nu_0}T\partial_{T}\Lambda^{(0)}, \quad c_{65}=\frac{\beta_{21}}{\nu_0}, \quad c_{66}=-\frac{3\Lambda^{(0)}}{2\nu_0}-\frac{1}{\nu_0}T\partial_T\Lambda^{(0)}+\frac{\beta_{22}}{\nu_0}
+\frac{3\omega^*\xi^{*1/3}}{M_2^\beta},
\eeq
\beq
\label{b20}
c_{70}=\left[-\frac{\overline m\epsilon_{12}}{T\nu_0}+(d+2)\frac{\xi^*}{2M_1^\lambda}+(d+2)\frac{\omega^*\xi^{*1/3}}{2}\frac{\delta m_\beta \overline mx_1\chi_1}{x_1m_1+x_2m_2}\right]D_U^*+(d+2)\frac{\omega^*\xi^{*1/3}}{2}\frac{\delta m_\beta m_2x_1x_2\chi_1}{x_1m_1+x_2m_2},
\eeq
\beq
\label{b21}
c_{77}=-\frac{\Lambda^{(0)}}{\nu_0}+\frac{\beta_{11}}{\nu_0}+\frac{3\omega^*\xi^{*1/3}}{M_1^\beta}, \quad c_{78}=\frac{\beta_{12}}{\nu_0},
\eeq
\beq
\label{b22}
c_{80}=\left[\frac{\overline m\epsilon_{21}}{T\nu_0}-(d+2)\frac{\xi^*}{2M_2^\lambda}-(d+2)\frac{\omega^*\xi^{*1/3}}{2}\frac{\delta m_\beta \overline mx_2\chi_2}{x_1m_1+x_2m_2}\right]D_U^*-(d+2)\frac{\omega^*\xi^{*1/3}}{2}\frac{\delta m_\beta m_1x_1x_2\chi_2}{x_1m_1+x_2m_2},
\eeq
\beq
\label{b23}
c_{87}=\frac{\beta_{21}}{\nu_0}, \quad c_{88}=-\frac{\Lambda^{(0)}}{\nu_0}+\frac{\beta_{22}}{\nu_0}+\frac{3\omega^*\xi^{*1/3}}{M_2^\beta}.
\eeq
In equations \eqref{b8}--\eqref{b23}, the fourth cumulants $K_i$ are defined by Eq.\ \eqref{3.14} and we have introduced the quantities
\beqa
\label{b24}
\nonumber \beta_{11}&=&-\frac{\nu_{11}}{4}\frac{(1+\alpha_{11})}{d(d+2)}\left[\alpha_{11}(d+8)-5d-4\right] \\ && -\nu_{12}\mu_{21}\frac{(1+\alpha_{12})}{d(d+2)}\Big\{\mu_{21}(1+\alpha_{12})\left[d+8-3\mu_{21}(1+\alpha_{12})\right]-3(d+2)
\Big\},  \\
\label{b25}
\beta_{12}&=&-3\nu_{12}\mu_{21}^3\frac{(1+\alpha_{12})^3}{d(d+2)}\frac{\rho_1}{\rho_2},
\eeqa
\beqa
\label{b26}
\epsilon_{12}&=& -\frac{\nu_{11}}{8}\frac{(1+\alpha_{11})}{d(d+2)}\left[\alpha_{11}(d^2-2d-8)+3d(d+2)\right]\frac{T_1^{(0)}}{m_1} -\frac{\nu_{12}}{2}\mu_{21}\frac{(1+\alpha_{12})}{d}\Bigg\{\mu_{21}(1+\alpha_{12})\nonumber\\
& & \times
\Big[d-3\mu_{21}(1+\alpha_{12})+2\Big]\frac{T_2^{(0)}}{m_2} -\frac{x_1}{x_2}\Big[d+3\mu_{21}^2(1+\alpha_{12})^2-6\mu_{21}(1+\alpha_{12})+2\Big]\frac{T_1^{(0)}}{m_1}\Bigg\}.
\eeqa
The expressions of $\beta_{22}$, $\beta_{21}$, and $\epsilon_{21}$ can be easily inferred from the forms of $\beta_{11}$, $\beta_{12}$, and $\epsilon_{12}$, respectively, by interchanging $1\leftrightarrow 2$.

%%%%%%%%%%%%%%%%%%%%%%%%%%%%%%%%%%%%%%%%%%%%%%%%%%%%%%%%%%%%%%%%%%%%%%%%%%%%%%%%%%%%%%%%%%%%%%%%%%

\bibliography{Maxwell}

\end{document}